\documentclass[12pt]{article}
\usepackage{a4wide,amsmath,amssymb,bm,graphicx}
\usepackage{jheppub}
\usepackage{slashed}
\usepackage{cancel}
\usepackage{pstricks}

\usepackage[nodisplayskipstretch]{setspace}
\numberwithin{equation}{section}

\newcommand{\defas}{\mathrel{\mathop:}=}

\newcommand{\tSih}{{\Sigma_h}}

\newcommand{\tPi}{\tilde{\Pi}}
\newcommand{\tD}{\tilde{D}}
\newlength{\blength}
\newlength{\slength}
\newlength{\hlength}
\newcommand{\myhl}{\vspace{\hlength}}

\newcommand{\ice}[1]{{}}
\newcommand{\EQN}[1]{\hspace{3mm}\fbox{\fbox{$#1$}} \label{#1}}
\newcommand{\mlabel}[1]{ \text{\footnotesize \hspace{3mm}\fbox{\fbox{$#1$}} \label{#1}}}
\renewcommand{\EQN}[1]{\mlabel{#1}}
\renewcommand{\mlabel}[1]{\label{#1}}
\renewcommand{\EQN}[1]{\label{#1}}

\newcommand{\nnb}{\nonumber}
\newcommand{\ed}{\end{document}}
\newcommand{\prd}{\partial}
\newcommand{\ep}{\epsilon}
\newcommand{\beq}{\begin{equation}}
\newcommand{\eeq}{\end{equation}}
\newcommand{\bea}{\begin{eqnarray}}
\newcommand{\eea}{\end{eqnarray}}
\newcommand{\bal}{\[\begin{aligned}}
\newcommand{\eal}{\end{aligned}\]}
\newcommand{\eqs}[1]{\begin{align}#1\end{align}}

\newcommand{\re}[1]{(\ref{#1})}

\newcommand{\g}{\gamma}
\newcommand{\ga}{\gamma}
\newcommand{\Ga}{\Gamma}
\newcommand{\al}{\alpha}

\newcommand{\Si}{\Sigma}
\newcommand{\om}{\omega}
\newcommand{\bc}{\begin{center}}
\newcommand{\ec}{\end{center}}
\newcounter{VBQ}

\newcommand{\vv}{v^2}
\newcommand{\Q}{\mathbb{Q}}
\newcommand{\F}{\mathbb{G}}
\newcommand{\jj}{Q}
\newcommand{\G}{G}
\newcommand{\TT}{{\perp\perp}}
\newcommand{\PT}{{\parallel\perp}}
\newcommand{\NEQN}[1]{\nnb}
\title{
Renormalization of parton quasi-distributions beyond the leading order:
spacelike vs.\ timelike\\
}
\author[a]{V. M.  Braun,}
\author[b,c]{K. G. Chetyrkin,}
\author[c]{B. A. Kniehl}
\affiliation[a]{
   Institut f\"ur Theoretische Physik, Universit\"at
   Regensburg, 93040~Regensburg, Germany}
\affiliation[b]{Institut f\"ur Theoretische Teilchenphysik, Karlsruhe
  Institute of Technology (KIT), Wolfgang-Gaede-Stra\ss{}e~1, 76131~Karlsruhe, Germany}
\affiliation[c]{
II. Institut f\"ur Theoretische Physik,
Universit\"at  Hamburg, Luruper Chaussee~149, 22761~Hamburg, Germany
}
\emailAdd{vladimir.braun@ur.de}
\emailAdd{Konstantin.Chetyrkin@kit.edu}
\emailAdd{kniehl@desy.de}
\abstract{
 We argue that the renormalization factors for nonlocal quark-antiquark 
 and gluon
 operators at space-like and time-like separations connected by a Wilson line
 coincide to all orders in perturbation theory.
 We calculate the anomalous dimensions and renormalization constants 
 of quark-antiquark and gluon operators to three- and two-loop accuracy, respectively,
 and also compute vacuum expectation values of these operators to three-loop accuracy.
}

\begin{document}
\maketitle
\renewcommand{\thefootnote}{\fnsymbol{footnote}}
\section{Introduction}
Studies of non-local gauge-invariant operators containing segments of Wilson lines have a long history.
They emerged in connection with the attempts to reformulate gauge theories in terms
of path-ordered gauge factors (Wilson lines), in particular within the loop-space formalism 
by Makeenko and Migdal \cite{Makeenko:1980vm}.   
The study of the renormalization of Wilson lines has been initiated by
Polyakov \cite{Polyakov:1980ca},
Gerwais and Neveau \cite{Gervais:1979fv}, and continued by several authors 
\cite{Dotsenko:1979wb,Craigie:1980qs,Arefeva:1980zd,Brandt:1981kf}; see Ref.~\cite{Dorn:1986dt} for a review.
The one-dimensional auxiliary-field formalism introduced in this context 
in Refs.~\cite{Gervais:1979fv,Arefeva:1980zd} enables the application of the usual language 
of correlation functions of local operators and is important both conceptually and at a 
technical level, as a basis for multiloop calculations.  
Subsequent applications of these methods have been to the study of infrared singularities 
in Feynman amplitudes, starting from the work of Refs.~\cite{Korchemsky:1985xj,Korchemsky:1987wg}, heavy-quark
effective theory (HQET) \cite{Korchemsky:1991zp,Broadhurst:1991fz,Chetyrkin:2003vi,Grozin:2007ap,Grozin:2008nu} and
transverse-momentum-dependent (TMD) factorization~\cite{Collins:2011zzd}.
  
Recently, there has been renewed interest in the study of matrix elements of non-local off-light-cone 
operators of the type
\begin{align}
 \Q(z) &= \bar{q}(z v) \,\Gamma [zv,0] \, q(0)\,, 
\mlabel{O(z)}
\\
 \F_{\mu\nu\alpha\beta}(z) &=  g^2 F_{\mu\nu}(z v) \,[zv,0] \, F_{\alpha\beta}(0)\,, 
\mlabel{F(z)}
\end{align}
where $q(x)$ is a quark field, $F_{\mu\nu}(x)$ is the gluon field strength tensor, $\Gamma$ is a certain Dirac structure, 
$v^\mu$ is an auxiliary four-vector and $z$ is a real number.
In addition, $[zv,0]$ is a straight-line-ordered Wilson line connecting the two fields,
\begin{align}
 [zv,0] &=  {\cal P} \exp \left[
    ig\, \int_0^z \! d z^\prime\, v^\mu A_{\mu}(z^\prime  v ) \right], 
\mlabel{wline}
\end{align}
which we assume to be taken in the proper representation of the gauge group,
fundamental for quarks and adjoint for gluons.  
Matrix elements of such operators acting on hadron states with large momenta are often referred to as 
parton quasi-distributions (qPDFs) \cite{Ji:2013dva} or pseudo-distributions~\cite{Radyushkin:2017cyf}.  
They can be factorized in terms of parton distribution functions (PDFs)~\cite{Izubuchi:2018srq} 
and, at the same time, are accessible in lattice calculations if the quark-antiquark separation 
is chosen to be space-like,  $ \vv \equiv v^\mu v_\mu <0 $. 
It was suggested that PDFs can be constrained in this way~\cite{Ji:2013dva}, and this possibility 
is being intensively explored, see, e.g., Refs.~\cite{Lin:2017snn,Cichy:2018mum} for reviews.    
The rationale for using matrix elements of the operators in Eqs.~\eqref{O(z)}
and \eqref{F(z)} in these studies is that 
they are ``cheaper'' to compute on the lattice as compared to other Euclidean observables with similar 
factorization properties; see, e.g., Refs.~\cite{Detmold:2005gg,Braun:2007wv,Ma:2017pxb}. This 
advantage comes at the price that the renormalization of the non-local operators in Eqs.~\eqref{O(z)} and \eqref{F(z)} is 
nontrivial and requires special 
attention~\cite{Xiong:2013bka,Ji:2015jwa,Ji:2017oey,Ishikawa:2017faj,Wang:2017eel,Wang:2017qyg,Izubuchi:2018srq,Zhang:2018diq,Wang:2019tgg,Balitsky:2019krf}.

In this paper, we address the question whether computational methods familiar
from HQET can be applied to the calculation of the renormalization constants
(RCs) of the operators in Eq.~\eqref{O(z)}, alias for qPDFs, in high
orders. The difference is that, in HQET, the heavy-quark velocity $v^\mu$ is
time-like, $\vv>0$, while, for qPDF studies, it is space-like, $\vv <0$.  We
argue that the change of sign has no effect on the renormalization and confirm
this result by an explicit calculation to three-loop accuracy. This result is
also relevant in the context of TMD factorization, where Wilson lines are
shifted off the light cone to regularize rapidity divergences in TMD
operators~\cite{Collins:2011zzd}. Our statement is that the anomalous
dimensions (ADs) and RCs do not depend on the direction, space-like or
time-like.
To avoid confusion, in this work, we imply using dimensional regularization.
In renormalization schemes with an explicit regularization scale, the Wilson
line in Eq.~\eqref{wline} suffers from an additional linear ultraviolet
divergence~\cite{Polyakov:1980ca}, which has to be removed.  This can be done
by mass renormalization, similarly to the introduction of the residual mass
term in HQET, or, alternatively, by considering a suitable ratio of matrix
elements involving the same
operator~\cite{Orginos:2017kos,Braun:2018brg}. Having in mind the second
approach, we calculate in this work the vacuum expectation values (VEVs) of
the operators in Eqs.~\eqref{O(z)} and \eqref{F(z)} to three-loop accuracy
for space-like and time-like choices of the auxiliary vector $v^\mu$.  VEVs of
nonlocal gluon operators are also of interest for studies of the QCD vacuum
structure; see Ref.~\cite{DiGiacomo:2000irz} for a review and further references.
The gluon case is also interesting as the generic gluon operator as defined in
Eq.~\eqref{F(z)} is not renormalized multiplicatively.  The calculation of its VEV
allows one to obtain the renormalization constants avoiding the necessity to
consider mixing with non-gauge-invariant operators. In this way, we verify the
mixing pattern found in Refs.~\cite{Dorn:1980hs,Dorn:1981wa} and also calculate the two-loop
mixing matrix, which is another new result.

This paper is organized as follows.
In Section~\ref{sec:two}, we recall the standard argument that any Green function in QCD with the
insertion of the non-local operator in Eq.~\eqref{O(z)} may be obtained from the
correlation function of two appropriate local ``heavy-light'' operators and
verify this diagrammatically.
In Section~\ref{sec:three}, we introduce the formalism to be used here, explain
the reduction of Feynman diagrams to master integrals and discuss the relation
of Green functions between the time-like and space-like regions.
In Section~\ref{sec:four}, we explain how the relationship between time-like
and space-like Green functions translates from momentum space to position
space.
In Section~\ref{sec:five}, we present our analytic and numerical results for
the correlators of interest here, both in momentum and position space.
Section~\ref{sec:six} contains our conclusions.
For the reader's convenience, we list the analytic results through three loops
for the ADs and RCs entering our analysis in Appendices~\ref{app:a} and
\ref{app:b}, respectively.
\section{Effective field theory}
\label{sec:two}
As is well known, the interaction of a particle propagating along a classical path in the background gauge field
reduces to the path-ordered phase factor along its trajectory. 
Such an auxiliary classical particle can be simulated by supplementing the QCD Lagrangian $\mathcal{L}_\mathrm{QCD}$
by an extra term,
\begin{align}
\mathcal{L}   &= \mathcal{L}_\mathrm{QCD} +\bar{h}_v\, \mathrm{i} v^{\mu}\mathcal{D}_{\mu} \, h_v\,,
\mlabel{lgr}
\end{align}
where $h_v$ is a (complex) scalar field in either fundamental or adjoint
representation of the gauge group, 
$\mathcal{D} = \prd_{\mu} -i g A^a_{\mu} T^a$ is the covariant derivative
and
$T^a$ are the  SU(3) generators in the appropriate representation. The Lagrangian in Eq.~\eqref{lgr} for
 $\vv >0$ is, essentially, the standard 
 HQET Lagrangian~\cite{Neubert:1993mb}, apart from our choice of $h_v(x)$ as
 a scalar. The case $\vv  < 0$   is of interest in connection with qPDFs
\cite{Ji:2017oey, Zhang:2018diq}.
Without loss of generality, we can assume $v_0>0$ and the usual causal boundary conditions for the
``heavy'' field $h_v$. Its free propagator reads
\begin{align}
  S_h^{(0)}(x) \equiv  \langle 0 |\text{T}\,\{h_v^i(x) \bar h_v^j(0)\} |0\rangle &=
\delta^{ij} \int\frac{ d^Dk}{i\,(2\pi)^D }\, e^{-ik\cdot x} \frac{1}{-v\cdot k-i\epsilon}
\nnb \\
&= 
\delta^{ij}\int_0^\infty \!\! ds \, \delta^{(D)}(x - sv) 
 \nnb \\
&= \delta^{ij}\,\frac{1}{|v|}\theta\left(\frac{v\cdot x}{\vv}\right)\,\delta^{(D-1)}(x_\perp)\,, 
       \mlabel{sprop}
\end{align}
where  $x_\perp^\mu = x^\mu - v^\mu (v \cdot x)/\vv $ and $i,j$ are color indices. 
Adding the interactions with the gluon field, one obtains
\cite{Dorn:1986dt}\footnote{%
Notice that the straight-line-ordered Wilson line $[x,0]$
is the unique solution of the differential equation $(x\cdot D)[x,0]=0$ with the 
boundary condition $[0,0]=1$, whereas the propagator of the ``heavy'' field is a Green function of the same
operator. Thus they differ by a factor which is just the free propagator.}
\begin{align}
  \langle 0 |\text{T}\,\{h_v^i(x) \bar h_v^j(0)\} |0\rangle_A &= S_h^{(0)}(x) [x,0]\,,
\end{align}
so that any Green function in QCD with the insertion of the non-local operator in Eq.~\eqref{O(z)}
can equivalently be obtained from the correlation function of two local ``heavy-light'' operators.
E.g. for quarks, 
\begin{align}
 S_h^{(0)}(vz)\langle 0| \text{T}\,\{O(z)\, \Phi(x_1,\ldots,x_n)\} |0\rangle = \langle 0|
\text{T}\,\{ \big(\bar{q} h_v\big)(z v) \Gamma \big(\bar{h}_vq\big)(0)\,\Phi(x_1,\ldots,x_n)\} |0\rangle\,,
\mlabel{equality}
\end{align}
where $\Phi(x_1,\ldots,x_n)$ stands for an arbitrary set of QCD fields at positions $x_1,\ldots,x_n$,
and similarly for gluons.
With our choice $v_0>0$ and for $z>0$, the path ordering in the operator on the l.h.s.\ of Eq.~\eqref{equality}
is consistent with time ordering. For $z<0$, the expressions on the l.h.s.\ and r.h.s.\ of Eq.~\eqref{equality} both vanish.
\begin{figure}[t]
\begin{center}
\includegraphics[angle=0,width=0.85\textwidth]{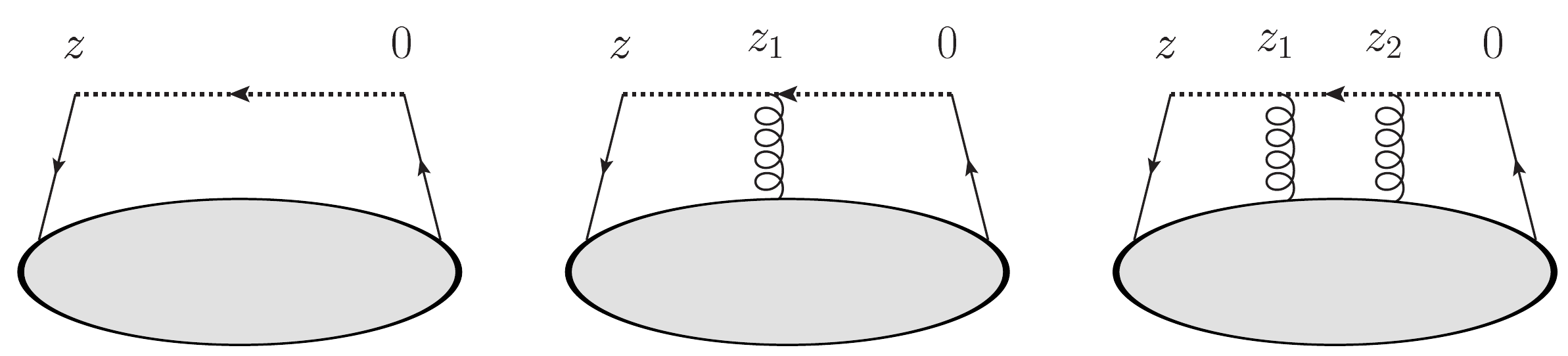}
\end{center}
\caption{\small The first three ``generic'' diagrams for an insertion of the operator $\Q(z)$ in Eq.~\eqref{O(z)}.
Wilson lines are indicated as dotted.}
\mlabel{fig:1}
\end{figure}
It is instructive to verify the equivalence in Eq.~\eqref{equality} diagrammatically at the few lowest orders 
in perturbation theory. For definiteness, consider quark operators. 
On the one hand, from the definition of the path-ordered
exponential in Eq.~\eqref{wline}, one obtains
\begin{align}
 \Q(z) &= \bar{q}(zv) \Gamma\left[
1+ ig \int_0^z\! d z_1 A_v (z_1v) + (ig)^2 \int_0^z\! d z_1\! \int_0^{z_1} d z_2 A_v (z_1v) A_v (z_2v) + \cdots\right] q(0)\,,
\mlabel{wl:3terms-a}
\end{align}
where we have used the shorthand notation $A_v = v^\mu A_\mu$. Notice that the factor $1/2!$ that naively appears in the third term
of the expansion of the exponential function is replaced by the ordering of the integration regions.
On the other hand, one can easily write the corresponding expression using the Feynman rules 
of the effective theory in Eq.~\eqref{lgr},
\begin{eqnarray}
\lefteqn{\hspace{-1cm}\text{T}\,\left\{\left(\bar{q} h_v\right)(z v)\, \Gamma \left(\bar{h}_vq\right)(0)\right\} =
\frac{1}{|v|} \theta(z)\delta^{(D-1)}(0_\perp)\bar q(zv)\Gamma\left[
  1 + ig \int_0^\infty\!\! d z_1\,\theta(z-z_1) A_v (z_1v)
  \right.}
  \nonumber\\
&&{}+ (ig)^2 \int_0^\infty\!\!d z_2\!\int_0^\infty\!\!d z_1\, \theta(z-z_1-z_2) A_v(z_1v+z_2 v) A_v(z_2 v)+\ldots\biggr] q(0)\,. 
\mlabel{wl:3terms-b}
\end{eqnarray}
In this way, the $1/2!$ factor in the last term is compensated by a different
mechanism: there are two equivalent ways to couple the four $h_v$ fields in
the Lagrangian insertions $ (ig)^2/2! \left(\bar{h}_v A_v^a T_a h_v \right)^2$
with those in the currents $\bar{q} h_v (z v)$ and $\bar{h}_vq(0)$.
Notice that disconnected diagrams are not counted.  The two expressions in
Eqs.~\eqref{wl:3terms-a} and \eqref{wl:3terms-b} are obviously equal up to an
overall factor, which is nothing but the free propagator of the ``heavy''
field, $S_h^{(0)}(zv)$.
\section{Calculation}
\label{sec:three}
\subsection{Generalities}
\newcommand{\sB}{\scalebox{.55}{$B$}}
We define the renormalization factors for generic composite operators  $O$ as
\begin{align}
    O(z) = Z_O O_{\sB}\,,    
\mlabel{def:Z}
\end{align}
where $O_B$ is the corresponding bare operator and  the bare fields are related to the 
renormalized ones as
\beq
A_{\mu}^{a, \sB} = \sqrt{Z_3}\,A_{\mu}^{a}\,, \qquad 
q_{\sB}  =  \sqrt{Z_2}\,  q\,, \qquad
h_v^{\sB} = \sqrt{Z_h} \, h_v\,.
\mlabel{RfromB:fields}
\eeq
The bare QCD coupling constant $g_{\sB}$ is expressed as
\beq
g_{\sB}   = Z_g \mu^{\ep}g\,,
\eeq
where $\mu$ is  the {}'t~Hooft mass and $\ep =(4-D)/2$, with  $D$ being
the running space-time dimension within  the method of dimensional regularization
\cite{Cicuta:1972jf,Ashmore:1972uj,tHooft:1972fi}.
Within the modified minimal-subtraction ($\overline{\mbox{MS}}$) scheme, every
RC is independent of dimensional parameters (masses and momenta) and can
be represented as
 \begin{eqnarray} 
Z(a) = 1 + \sum_{n=1}^\infty \frac{z^{(n)}(a)}{ \epsilon ^n}
\mlabel{}
\,,
 \end{eqnarray} 
where $a = g^2/(16 \pi^2)$.
Given a RC $Z(a)$,  the  corresponding AD is defined as
 \begin{equation} 
\gamma(a) = \pm \mu^2\frac{d \ln Z(a)}{d \mu^2}
=\pm \left( -  a \frac{ \partial  z^{(1)}(a)}{ \partial  a}\right)
=  \sum_{n=1}^\infty (\gamma)_n \, a^{n} 
\mlabel{anom:dim:generic}
\,.
\end{equation}
Here, the sign depends on the way  how the considered bare  quantity, $O_B$, is related  to its
renormalized counterpart, $O$. 
With our convention \eqref{def:Z},
the renormalization group (RG) equation for $O$ assumes the form
\beq
\mu^2\frac{d O}{d \mu^2}  = \ga_O O \qquad \text{with} \qquad 
\ga_O = \mu^2\frac{d \ln Z_O(a)}{d \mu^2}
\,.
\eeq
Traditionally, in renormalizing coupling constants, wave functions and masses, the
defining  relation~\re{def:Z} is written with $Z_O$ on the left-hand side, e.g.,
\[
  Z_m\, m = m_B {}.
\]  
 As a result,  the ADs $\ga_2$ and $\gamma_h$, and the $\beta$ function
(see below) are defined  with the minus sign  version of Eq.~\re{anom:dim:generic},
but the ADs of composite operators (see below) 
with the plus sign.
Customarily, one also  defines $Z_a =Z_g^2$ and refers to the corresponding AD
as the QCD $\beta$ function,
 \begin{equation} 
\beta(a) =
a\, \frac{ \partial  z^{(1)}_a(a)}{ \partial  a}
=
 \sum_{n=1}^\infty \beta_{n} a^{n}
\mlabel{beta:function:generic}
\,.
 \end{equation} 
The RCs $Z_2 $ and $Z_a$ serve to renormalize the
standard QCD Lagrangian and have been well known through three loops for a long time
\cite{Tarasov:1980au,Tarasov:2019rwk,Larin:1993tp}.
The RC $Z_h$ is  known to the same accuracy for the time-like case 
$\vv >  0$ \cite{Chetyrkin:2003vi} and will be used by us as convenient reference point. 
For the reader's convenience, we quote the RCs as well as the corresponding ADs
in Appendices~\ref{app:b} and \ref{app:a}, respectively.
Due to an overall $\delta^{(D-1)}(x_{\perp})$ factor in position space,
the self-energy and hence also the propagator of the ``heavy'' field $h_v$ 
can only depend on the projection of its momentum onto $v^\mu$, for which we use the notation
\begin{align}
   \omega = p\cdot v\,.
\end{align} 
The heavy-field  self-energy is defined, as usual, as the sum of one-particle-irreducible amputated diagrams, 
\begin{align}
\omega\, \Si_h(\omega) = i\int\! d^Dx\,  e^{ip\cdot x}
\langle 0|T\{ h_v(x)\  \bar{h}_v(0) \}|0\rangle^{\text{1PI, amputated}}\,,
\EQN{sigma:def}
\end{align}
and the full propagator is then given by
\begin{align}
  \tilde{S}_h(\omega) &= \frac{\tilde{S}^0_h(\omega)}{1+\Si_h(\omega)}\,, \qquad \tilde{S}^0_h(\omega) = \frac{1}{-\omega -i0}\,.
\mlabel{S(omega)}
\end{align}
Notice that the full propagator and the 
self-energy in Landau gauge satisfy the RG equations,
%
%
\begin{eqnarray}
\Bigl(\mu^2\frac{\prd }{\prd \mu^2} + \beta(a)\, a\, \frac{\prd }{\prd a}\Bigr) \tilde{S}_h &=& \ga_h \,  \tilde{S}_h\,,\\
\Bigl(\mu^2\frac{\prd }{\prd \mu^2} + \beta(a)\, a\, \frac{\prd }{\prd a}\Bigr) {\Sigma}_h &=& -\ga_h \, {\Sigma}_h\,.
\EQN{RG:hself}
\end{eqnarray}
 
The renormalization factors $Z_Q$ of the ``heavy-light'' operators in
Eq.~\re{current} 
are calculated from the 
corresponding vertex functions $\Gamma(p,\omega)$ as explained, e.g., in
Ref.~\cite{Grozin:2004yc};
the general formalism is the same for $\vv>0$ and $\vv<0$.

\subsection{Renormalization constants and VEVs of qPDF operators} 

Let us first consider quark operators.
Let 
\begin{align}
\mlabel{current}
  \jj_i(x) = \bar q_i(x) h_v(x)\,,
\end{align}
where $i$ is the spinor index, which we do not show in what follows.
One can show that this operator is multiplicatively renormalized. 
The corresponding RC $Z_{\jj}$,  
\begin{align}
    \jj(x) = Z_\jj \jj_{\sB}(x)\,, 
\mlabel{BfromR:ops}   
\end{align}
was calculated to three-loop accuracy for the time-like case $\vv >  0$ \cite{Chetyrkin:2003vi}
and is the same for space-like $\vv <  0$. 
The RC $Z_{\Q}$ of the nonlocal operator defined in Eq.~\eqref{O(z)} is related to $Z_{\jj}$
as~\cite{Ji:2017oey}
\begin{align}
 \Q(z) = Z_{\Q} \Q_{\sB}(z)\,, \qquad  Z_{\Q} = Z_{\jj}^2.
\EQN{ZQZh}
\end{align} 
The corresponding ADs are obviously related as
\begin{align}
 \ga_\Q(z) = 2\,\ga_Q.
\EQN{gaQgah}
\end{align}
 The  equality \re{ZQZh} implies that $Z_{\Q}$ does not depend on the Dirac structure
$\Gamma$ in the definition of the operator $\Q(z)$, Eq.~\eqref{O(z)}.  We have
checked this relation at the two-loop level by explicit calculation of the VEV of
$\Q(z)$ to three-loop accuracy; see Section~\ref{sec:five}.
This VEV is non-vanishing in perturbation theory only for 
the Dirac structure $\Gamma = \gamma_\mu$, in which case it follows from Lorentz invariance that 
$\langle 0|\Q_\mu|0\rangle \propto v_\mu$. Thus, it is sufficient to consider 
\begin{align}
  \Pi(z) =   \langle 0|v^\mu \Q_\mu(z) |0\rangle =  \langle 0| \bar q(zv)\slashed{v}[zv,0] q(0)|0\rangle\,,   
\mlabel{<O(z)>}
\end{align}  
or, equivalently, the momentum space correlation function,
\begin{align}
  \widetilde{\Pi}(\omega) & =  i\int\!d^D x \, e^{ip\cdot x} \langle 0| \text{T}\,\{\jj(x) \slashed{v} \jj^\dagger (0)\}  
\mlabel{Pi(omega)}\rangle\,,
\end{align}
where the current $\jj(x)$ is defined in Eq.~\eqref{current}.
It is easy to see that
\beq
\Pi(z > 0) =  \frac{1}{i} 
\int d^{D-1}x_{\perp}      
\int \frac{d^Dp}{(2\pi)^D}  e^{-ip\cdot x} \tilde{\Pi}(p)
 \qquad \text{with} \qquad  z= \pm x\cdot v\,,
\eeq
where the $\pm$ sign corresponds to the choice $\vv = \pm 1$.
Notice that $\Pi(z > 0 )$ does not depend on  the ``contact'' terms in 
$\tilde{\Pi}(p)$ of the form $\mbox{\it const} \cdot\om^2$, which suffer 
from an extra UV divergence coming from the integration region around $x=0$ 
in Eq.~\re{Pi(omega)}.
Thus, in momentum space, it is
natural to consider instead of $\tilde{\Pi}(p)$ an analog of the Adler function, namely
\beq
\tD(\om) =  \om \frac{d}{d \om} \,\Bigl(  \frac{\pi^2}{\om^2} \tPi(\om) \Bigr)
{}.
\EQN{Adler}
\eeq
The corresponding RG equation for $D(\om)$ reads
  \beq
\Bigl(\mu^2\frac{\prd }{\prd \mu^2} + \beta(a)\, a\, \frac{\prd }{\prd a}\Bigr)\, \tD = 2\ga_Q \,\tD 
{}.
\EQN{RG:D}
\eeq
The gluon case is more involved because, in the presence of the external four-vector $v^\mu$, the components parallel 
and transverse to  $v^\mu$ can be renormalized differently. Introducing the corresponding projection operators,
\begin{align}
   g_{\mu\nu}^\parallel = \frac{v_\mu v_\nu}{v^2}\,, \qquad
   g_{\mu\nu}^\perp = g_{\mu\nu} - \frac{v_\mu v_\nu}{v^2}\,,  
\end{align}
we can define two multiplicatively renormalizable gauge-invariant local operators in effective theory with adjoint ``heavy'' scalars as%
\footnote{Notice that we include the QCD coupling $g$ in the definition of the operators, which simplifies the renormalization factors.}
\begin{align}
 \G^{\parallel\perp}_{\mu\nu}(x) &=    
\Big[g^\parallel_{\mu\alpha}  g^\perp_{\nu\beta} - g^\parallel_{\nu\alpha}  g^\perp_{\mu\beta} \Big] g F^{\alpha\beta}(x) h_v(x)\,,
\notag\\
 \G^{\perp\perp}_{\mu\nu}(x) &=    g^\perp_{\mu\alpha}  g^\perp_{\nu\beta} g F^{\alpha\beta}(x) h_v(x)\,, 
\end{align}
with the RCs $Z_{\parallel\perp}$ and $Z_{\perp\perp}$,
\begin{align}
    \G^{\parallel\perp}_{\mu\nu} = Z_{\parallel\perp} \big(\G^{\parallel\perp}_{\mu\nu}\big)_B\,, \qquad
    \G^{\perp\perp}_{\mu\nu} = Z_{\perp\perp} \big(\G^{\perp\perp}_{\mu\nu}\big)_B\,.
\mlabel{RC-gluons}
\end{align}
We will denote the corresponding ADs as $\ga_\PT$ and $\ga_\TT$ respectively. 
The correlation functions of these operators have the form
\begin{align}
 \langle 0| G^{\perp\perp}_{\mu\nu}(x)  \bar G^{\perp\perp}_{\alpha\beta}(0) |0 \rangle 
&=  (g^\perp_{\mu\alpha}g^\perp_{\nu\beta} - g^\perp_{\nu\alpha}g^\perp_{\mu\beta}) \Pi_{\perp\perp}(x) \,,
\notag\\
 \langle 0| G^{\parallel\perp}_{\mu\nu}(x)  \bar G^{\parallel\perp}_{\alpha\beta}(0) |0 \rangle
&= (g^\parallel_{\mu\alpha} g^\perp_{\nu\beta} -  g^\parallel_{\nu\alpha} g^\perp_{\mu\beta}
- g^\parallel_{\mu\beta} g^\perp_{\nu\alpha} +  g^\parallel_{\nu\beta} g^\perp_{\mu\alpha})\Pi_{\parallel\perp}(x) \,,
\notag\\
 \langle 0| G^{\perp\perp}_{\mu\nu}(x)  \bar G^{\parallel\perp}_{\alpha\beta}(0) |0 \rangle
&= 0\,,  
\end{align}
where $\bar  G^{\perp\perp}_{\alpha\beta} = g^\perp_{\mu\alpha}  g^\perp_{\nu\beta} g F^{\alpha\beta} \bar h_v$, etc., and are renormalized by squares of the corresponding RCs in Eq.~\eqref{RC-gluons}. 
In this notation, a generic two-gluon vacuum correlation function related to the qPDF operator in Eq.~\eqref{F(z)} 
takes the form
\begin{align}
 \Pi_{\mu\nu\alpha\beta}(x) &= \langle 0| g^2 F_{\mu\nu}(x) h_v(x) \bar h_v(0) F_{\alpha\beta}(0)|0\rangle
\notag\\&=
(g^\perp_{\mu\alpha}g^\perp_{\nu\beta} - g^\perp_{\nu\alpha}g^\perp_{\mu\beta}) \Pi_{\perp\perp}(x) 
+ (g^\parallel_{\mu\alpha} g^\perp_{\nu\beta} -  g^\parallel_{\nu\alpha} g^\perp_{\mu\beta}
- g^\parallel_{\mu\beta} g^\perp_{\nu\alpha} +  g^\parallel_{\nu\beta} g^\perp_{\mu\alpha})\Pi_{\parallel\perp}(x)  
\notag\\&=
(g_{\mu\alpha}g_{\nu\beta} - g_{\nu\alpha}g_{\mu\beta})\, \Pi_{\perp\perp}(x)
\notag\\&\quad
+  \frac{1}{v^2}( v_\mu v_\alpha g_{\nu\beta} -  v_\nu v_\alpha g_{\mu\beta}
-v_\mu v_\beta g_{\nu\alpha} +  v_\nu v_\beta g_{\mu\alpha})\, [\Pi_{\parallel\perp}(x) - \Pi_{\perp\perp}(x)]\,.   
\end{align}
For future reference, we introduce the corresponding ``Adler'' functions:
\begin{eqnarray}
\tD_\TT(\om) &=&  \om \frac{d}{d \om} \,\Biggl(  \frac{ \tPi_\TT(\om)}{\om^3} \Biggr)\,,
\nonumber\\
\tD_\PT(\om) &=&  \om \frac{d}{d \om} \,\Biggl(  \frac{ \tPi_\PT(\om)}{\om^3} \Biggr)\,,
\end{eqnarray}
where
\beq
  \widetilde{D}_{\perp\perp}(\om)   =  i\int\!d^D x \, e^{ip\cdot x} \Pi_{\perp\perp}(x),\qquad
\widetilde{D}_{\parallel\perp}(\om)  =  i\int\!d^D x \, e^{ip\cdot x} \Pi_{\parallel\perp}(x)\,.
\mlabel{mom:Pitt+pt)}
\eeq
The functions $\tD_\TT(\om)$ and $\tD_\PT(\om)$ satisfy the standard RG
equations like Eq.~\re{RG:D} with the ADs $(2\,\ga_\TT)$ and
$(2\, \ga_\PT)$ respectively.
In the qPDF literature, following Ref.~\cite{Dorn:1980hs,Dorn:1981wa}, 
one usually introduces a different operator basis,
\begin{align}
J_1^{\mu\nu} &= g\,F^{\mu\nu}h_v\,, \qquad
J_2^{\mu\nu} = \frac{v_\rho}{v^2} g \left(
  F^{\mu\rho}_a v^\nu -  F^{\nu\rho}_a v^\mu
\right) h_v^a\,,
\end{align}
so that, obviously,
\begin{align}
  J_2^{\mu\nu}(x) &= G^{\parallel\perp}_{\mu\nu} (x)\,,
\qquad
  J_1^{\mu\nu}(x) =  G^{\perp\perp}_{\mu\nu}(x) +  G^{\parallel\perp}_{\mu\nu}(x)\,. 
\end{align}
The operators $J_1$ and $J_2$ are renormalized by a triangular $2\times2$ mixing matrix $Z_{ik}$ such 
that $Z_{21}=0$ and $Z_{12} - Z_{22}+ Z_{11} = 0$ to all orders. 
These relations imply that
\beq
Z_{11} =  Z_{\perp\perp}\,,\qquad
Z_{22} =  Z_{\parallel\perp}\,.
\eeq
Notice that we ignore mixing with gauge-noninvariant operators as they do not
contribute to gauge-invariant observables.
With the main definitions at hand, we proceed to describe the calculation procedure.

\subsection{Reduction}

The generation of the Feynman diagrams and their reduction to master integrals (MIs)
have been done in the  standard way, using the programs
QGRAF \cite{QGRAF} and FIRE6 \cite{Smirnov:2019qkx}, respectively.\footnote{We have also used
  the Mathematica  program LiteRed~1.4  \cite{Lee:2012cn,Lee:2013mka}  and  the REDUCE package Grinder
  \cite{Grozin:2000jv}  for testing purposes and the identification of the MIs.} 
  
The reduction typically results in a sum of MIs with coefficients being  
rational functions of the space-time dimension $D$. In addition, every MI is multiplied 
by a factor of the form 
\begin{align}
(\vv)^{n_1}\, \omega^{n_2}\,,
\mlabel{fctr}
\end{align}
and color factors like $C_F$, $C_A$, etc. 
We have checked that our results for the time-like  case $\vv = 1$
are in full agreement with Ref.~\cite{Chetyrkin:2003vi}.
In our calculation, we have used the values of the relevant MIs given in
Ref.~\cite{Chetyrkin:2003vi}.
\subsection{Master integrals: time-like versus space-like \mlabel{MIs}}
The MIs for $\vv >0$, which we refer to as time-like, are analytic
functions of $\omega$. They are real for $\omega <0$ and have  
a branch cut at $\omega>0$.
On dimensional grounds, each MI has the form
\begin{equation} M(\omega < 0, \vv = 1; \mathbf{n}) = (-2\omega)^{d^M}\, M(\ep; \mathbf{n}) \,,\qquad
d^{M} = D\, L  - 2\sum_l n_l - \sum_h n_h \,,
\end{equation}
where $M(\ep; \mathbf{n})$ is a real function of the space-time dimension
$D$, $d^{M}$ is the dimension of the MI $M$, $L$ stands for the number of loops and the sums over $l$ and $h$ count the indices $n_l$ and $n_h$ of all usual QCD (``light'') and special (``heavy'') 
lines of the integral. 
The argument ${\mathbf n}$ stands for the collection of all indices. 
We assume that every MI is of scalar type, so that the corresponding 
integrand is given by a product of denominators involving 
``light'' (massless) and ``heavy'' propagators, possibly raised to certain (integer) powers (indices). 
The reduction to scalar MIs is certainly possible at the three-loop level; see, e.g., Refs.~\cite{Grozin:2000jv,Chetyrkin:2003vi}.
The restriction to the normalization $\vv =1$ can easily be relaxed.
Indeed, the ``heavy'' propagator in Eq.~\eqref{sprop} is a homogeneous function w.r.t.\ the
rescaling  $v^\mu \to \lambda v^\mu$. Thus, we have
$M(\lambda\omega,\lambda^2\vv;\mathbf{n}) = \lambda^{-d^M_v}
M(\omega,\vv;\mathbf{n})$,
where $d^M_v = - \sum_h n_h$ stands for the $v$ dimension of the integral, 
so that, for generic time-like $\vv>0$, we have   
\begin{align}
M(\omega < 0, \vv >0; \mathbf{n}) &= (-2\omega)^{d^M} \, (\vv)^{(-d^M + d^M_v)/2}  M(\ep; \mathbf{n})
\notag\\&=
 (-2\omega )^{d^M_v} \left(\frac{4\omega^2}{\vv}\right)^{(d^M - d^M_v)/2} M(\ep;\mathbf{n})\,. 
\mlabel{Mt.generic}
\end{align}
The result for $\omega>0$ is obtained by analytic continuation. In this way,
the MI acquires an imaginary part according to the usual causal prescription 
$\omega \to \omega +i0$, so that $4\omega^2 \to (-2\omega- i0)^2$. 
In order to calculate a MI for the space-like case $\vv<0$, 
it is useful to start from the so-called $\alpha$ representation of the time-like MI,\footnote{%
Within HQET,
this was considered in Ref.~\cite{Grozin:2003eg}. A general discussion can be found, e.g., in Refs.~\cite{Smirnov:2012gma,Lee:2013mka}.} 
\begin{align}
  M(\omega, \vv >0; \mathbf{n})&=
  \frac{\Gamma\left(\Sigma n -Ld/2\right)}{\prod_{\alpha}\Gamma\left(n_{\alpha}\right)}
  \int\prod_{\alpha}dz_{\alpha}z_{\alpha}^{n_{\alpha}-1}\delta\left(1-\Sigma z\right)
  \frac{(F-i0)^{Ld/2-\Sigma n}}{U^{\left(L+1\right)d/2-\Sigma n}}  \,,
\mlabel{alpha.para}
\end{align} 
where $\Sigma n=\sum_{\alpha}n_{\alpha}\equiv \sum_l n_l + \sum_h n_h $, $\Sigma z=\sum_{\alpha}z_{\alpha}$, and
$U$ and $F$ are homogeneous polynomials of degree $L$ and $L+1$ in the integration parameters
$z_{\alpha}$, respectively. The function $U$ does not depend on kinematic invariants, whereas the 
function $F$ can be written as
\begin{align}
F&=  -2\,\omega\,  T_p + \vv \,T_v
\mlabel{F}
\,,  
\end{align}
with polynomials $T_p$ and $T_v$ that only depend on the parameters
$z_{\alpha}$ 
and are defined to be positive in the integration region of Eq.~\eqref{alpha.para}.
Notice that, in Eq.~\eqref{alpha.para}, we do not assume that $\omega <0$, so that, for $\vv >0$, the MI acquires an
imaginary part at $\omega>0$ according to the Feynman prescription $F\mapsto F-i0$.
The crucial observation is that, if one \emph{simultaneously} changes the signs of $\vv$ and $\omega$, the MI receives an overall phase factor,
\begin{align}
    M(\omega > 0 , \vv < 0; \mathbf{n})& = e^{-i\pi(Ld/2-\Sigma n)} 
    M(- \omega, |\vv| ; \mathbf{n})
\notag\\&=
   e^{i\pi \epsilon L}\, (-1)^{\Sigma n} M(- \omega, |\vv| ; \mathbf{n})\,,
\end{align} 
as, by definition,
\begin{align}
(F - i0)^{\lambda} \defas  
 \begin{cases}
  F^{\lambda}      & \text{if $F>0$}\,,
  \\
  (-F)^{\lambda} \,   e^{- i\pi \lambda} &  \text{if $F<0$}\,.
  \end{cases}
\mlabel{x-i0}
 \end{align}
Thus, we obtain
\begin{align}
     M(\omega > 0 , \vv < 0; \mathbf{n})& = 
 e^{i\pi \epsilon L}\, (-1)^{\Sigma n} |2\omega|^{d^M} \, |\vv|^{(-d^M + d^M_v)/2}  M(\ep;\mathbf{n})
\notag\\&=
e^{i\pi \epsilon L}\,
(-1)^{(d^M_0 + d^M_v)/2}
|2\omega|^{d^M} 
|\vv|^{(-d^M + d^M_v)/2}  M(\ep;\mathbf{n})
{}
\notag\\&=
%
%
(- 2\omega)^{d^M_v}  \left(\frac{4\omega^2}{\vv} - i0\right)^{(d^M-d^M_v)/2} M(\ep;\mathbf{n})\,,
\end{align}
where $d_0^M  =d^M|_{\ep=0}$.
A generic Green function $G(\omega, \vv)$ with mass dimension $d^G$ and $v$ dimension $d^G_v$ is given by the 
sum of MIs multiplied by extra kinematic factors $(\vv)^{j_1}(-2\omega)^{j_2}$,
where 
$j_1$ and $j_2$ are integers that satisfy the obvious relations 
\begin{align}
j_2 + d^M = d^G\,, \qquad  2 j_1 + j_2  + d^M_v = d^G_v\,.
\end{align}
It is easy to see that
\begin{align}
 (\vv)^{j_1}(-2\omega)^{j_2} M(\omega < 0, \vv > 0; \bf{n}) &= 
(-2\omega )^{d^G_v} \left(\frac{4\omega^2}{\vv}\right)^{(d^G-d^G_v)/2} M(\ep;\mathbf{n})\,, 
\notag\\
 (\vv)^{j_1}(-2\omega)^{j_2} M(\omega > 0, \vv < 0; \bf{n}) &=
 (-2\omega )^{d^G_v} \left(\frac{4\omega^2}{\vv}-i0\right)^{(d^G-d^G_v)/2} M(\ep;\mathbf{n})\,, 
\end{align}
so that, upon this multiplication, the mass and $v$ dimensions of 
a particular MI are substituted by those of the Green function in question 
and are the same for the contributions of all MIs and for all Feynman diagrams.
As a consequence, going over from $\vv>0,\, \omega <0$ to $\vv <0,\, \omega >0$,  
the Green function acquires an overall phase factor, 
\begin{align}
   G(\omega > 0, \vv < 0) &= e^{i\pi(d^G+d^G_v)/2} G(-\omega, |\vv|)
\notag\\&
= e^{i\pi\epsilon L} (-1)^{(d_0^G+d^G_v)/2} G(- \omega, |\vv| ) \,, 
\end{align}
where $d_0^G  =d^G|_{\ep=0}$, which is the final result. 
Thus, we conclude the following:
\begin{itemize}
\item
 A generic Green function at $\vv <0$ can be obtained from the result at $\vv>0$
 by the (possible) global sign change $(-1)^{(d_0^G+d^G_v)/2}$,
which is the same to all orders of perturbation theory,
and the formal substitution 
\begin{align}
  (-2\omega)^{d_0^G}\ln (-2\omega-i0)|_{\vv=1} \to (2\omega)^{d_0^G}
  \left[\ln (2\omega +i0) -  i\frac{\pi}{2}\right]_{\vv=-1}
\,,
\mlabel{t->s}
\end{align}
where we assume $|\vv|=1$ and the Feynman causal prescription.
\item
Since, in minimal schemes, the RCs and ADs neither depend on  the global sign nor on the value of the external momentum, 
 they are not affected by the analytic continuation in $\vv$ and are the same for time-like ($\vv =1$) and space-like ($\vv=-1$) kinematics.
\end{itemize}
\section{Transition to  position space}
\label{sec:four}
Without loss of generality, we may assume $v^\mu = (1,0,0,0)$ and $v^\mu = (0,0,0,1)$ for the time-like and space-like
cases, respectively. For this choice, the variable $\omega$ is the energy, $\omega = p_0$, for the
time-like case and the $z$ component of the momentum up to a minus sign, $\omega = - p_z$, for the space-like case.
The relation between generic correlation functions established in the previous section, therefore, 
connects a $v^2=1$ Green function for negative energy with the corresponding $v^2=-1$ Green function with negative
momentum in $z$ direction. 
The corresponding position-space variables for these cases are obviously the
separation in time $t$ and distance $z$ of the quark and the antiquark in the operator in Eq.~\eqref{O(z)}.     
In the time-like case, the transition is performed with the  help of the generic formula
\eqs{
\int \frac{d \om }{2\pi}\,  e^{-it\, \om } \, (-2\omega  - i0)^{-2L\ep  -n }
\, &=\,
\frac{e^{i (2 L\ep +n) \pi/2}}{2\,\Ga(n + 2\, L\,\ep)}\,\,\theta(t)\Bigl(\frac{t}{2}\Bigr)^{2 L \ep +n-1}
\nnb
\\
&=
\frac{i}{2\,\Ga(n  + 2\, L\,\ep)}\,\theta(t)\,\Bigl(\frac{i\,t}{2}\Bigr) ^{2 L \ep +n-1}
\,,
\EQN{x:t}
}
where we assume $n$ to be integer.
 
Thus, renormalized time dependent correlation functions are expressed in terms of 
the following combination
\beq
\ln\frac{i\,t\, e^{\ga_E}}{2} \equiv \ln\frac{t}{2} + \ga_E +  i\frac{\pi}{2}
\EQN{comb1}
\,, 
\eeq
where Euler's constant $\gamma_E$  appears  naturally 
due to a universal factor,
\beq
\Ga(n  + 2\, L\,\ep) \equiv \Ga(1  + 2\, L\,\ep)\, \left(1+2 L \ep\right)_{(n-1)} \,,
\eeq
with $\left(1+2 L \ep\right)_{(n-1)}$ being the Pochhammer symbol. 
There is no $\ga_E$ in momentum space results.
The analogue of Eq.~\eqref{x:t} for the space-like case is
\eqs{
\int \frac{d p_z }{2\pi}\, e^{i z\, p_z } \, (-2 p_z  + i0)^{-2 L \ep -n}
\, &=\,
\frac{e^{-i (2 L\ep +n) \pi/2}}{2\,\Ga(n + 2\, L\,\ep)}\,\,\Bigl(\frac{z}{2}\Bigr) ^{2 L \ep +n-1}
\nnb
\\
&=
\frac{-i}{2\,\Ga(n + 2\, L\,\ep)}\,\,\left(\frac{-i\,z}{2}\right)^{2 L \ep +n -1}
\,.
\EQN{x:s}
}
Comparing Eqs.~\eqref{x:t} and \eqref{x:s}, we infer that the transition from a
time-like to a space-like renormalized correlation function in position space amounts to the
formal substitution
\beq
\ln\frac{it}{2} \to \ln\frac{z}{2}\,,
\mlabel{sub}
\eeq
up to a possible change of the global sign.
\ice{
\section{Temporary: One-loop results}
For clarity repeat definitions from \eqref{<O(z)>} and \eqref{Pi(omega)}: 
\begin{align}
  \langle 0|v^\mu O_\mu(z) |0\rangle &=  \langle 0| \bar q(zv)\slashed{v}[zv,0] q(0)|0\rangle\,,   
\notag\\
  \Pi(\omega) & =  i\int\!d^D x \, e^{ip\cdot x} \langle 0| \text{T}\,\{j(x) \slashed{v} j^\dagger (0)\}  
\end{align}
I get for position-space: 
\begin{align}
   \langle 0|v^\mu O_\mu(z) |0\rangle &=  \frac{2iN_c}{\pi^2 \vv\, z^3} 
\end{align}
Momentum space $\vv =1$:
\begin{align}
  \Pi(\omega) & =  \frac{N_c}{2\pi^2} \omega^2
\Big\{\frac{1}{\epsilon} -\gamma_E + \ln 4\pi  -2 \ln (-2\omega/\mu) + 2 \Big\}
\end{align}
Momentum space $\vv =-1 $:
\begin{align}
  \Pi(\omega) & =  - \frac{N_c}{2\pi^2} \omega^2
\Big\{\frac{1}{\epsilon} -\gamma_E + \ln 4\pi  -2 \ln (2\omega/\mu) + 2 + i\pi \Big\}
\end{align}
I think the sign change is due to the kinematic factor $v^2$; Defining $v$-dimension as the sum of powers of 
heavy lines we did not take into account that explicit $v$-factors can be present. 
%
%
}
\section{Results}
\label{sec:five}
In this section, we collect our results for the case of standard QCD with the
SU(3) gauge group and $n_f$ active quarks triplets. 
Notice that the results for the self-energy and the propagator of the ``heavy'' field, 
the ADs $\ga_2$ and $\ga_h$ as well as the corresponding RCs are gauge dependent.
The expressions below are given in Landau gauge, as it is  most relevant  for
lattice applications.
Full results for a generic gauge group and including the
gauge as well as the momentum/position dependence are appended in the arxiv
submission of this paper as auxiliary
files in computer readable format.
Many results given in the text and in the auxiliary files have originally been
obtained by other authors and have been included here for completeness.
In particular the AD of the heavy-light current $\ga_\jj$ was computed at one, two and three loops in
Refs.~\cite{Politzer:1988wp,Shifman:1986sm}, Refs.~\cite{Ji:1991pr,Broadhurst:1991fz}
and Ref.~\cite{Chetyrkin:2003vi}, respectively.
The AD $\ga_h$ of the ``heavy''  field  $h_v$ was computed at two and
three loops in Ref.~\cite{Broadhurst:1991fz} and
Refs.~\cite{Melnikov:2000zc,Chetyrkin:2003vi}, respectively, and, recently, 
at four loops in Refs.~\cite{Marquard:2018rwx,Bruser:2019auj}.
The RCs $Z_3$, $Z_2$ and $Z_a$ have been known through three loops for a long
time \cite{Tarasov:1980au,Tarasov:2019rwk,Larin:1993tp}.
The VEV $\Pi$ in Eq.~\eqref{O(z)} was computed at two loops in Ref.~\cite{Broadhurst:1991fc} and at
three loops in Ref.~\cite{Czarnecki:2001rh}.
Notice that all these results for the ADs $\ga_h, \ga_{\jj}$ and for the VEV $\Pi$ were obtained for 
the time-like choice of the vector $v^\mu$, with $v^2 =1$. 
Our contribution is to clarify the changes for the space-time choice  $v^2 = - 1$. 
We have also computed the VEV of the gluon off-light-cone operator in Eq.~\re{F(z)} in the three-loop approximation
as well the corresponding  anomalous dimensions at two loops. Our results are in agreement
with the two-loop VEV found in Ref.~\cite{Eidemuller:1997bb} and the one-loop ADs first computed in Refs.~\cite{Dorn:1980hs,Dorn:1981wa}.  
\subsection{Anomalous dimensions and renormalization constants}
As follows from Eq.~\re{MIs}, ADs and RCs are not sensitive to the sign choice of
$\vv$. \ice{
 All the calculations performed here completely reproduce the
corresponding results listed in Ref.~\cite{Chetyrkin:2003vi}.
}
The analytic results for the ADs $\beta$, $\gamma_Q$, $\gamma_h$, $\gamma_2$,
$\ga_\TT$ and $\ga_\PT$
are listed in Appendix~\ref{app:a}, and those for the RCs
$Z_a$, $Z_Q$, $Z_h$, $Z_2$, $Z_\TT$ and $Z_\PT$ in Appendix~\ref{app:b}.

\subsection{Correlation functions: momentum space}
Our result for the self-energy of the ``heavy'' field, $\Si_h$, defined in Eq.~\re{sigma:def} 
for the space-like case $\vv=-1$ can be written as\footnote{%
In what follows, we do not show  the trivial  dependence of the self-energy $\Si_h(p)$ and
the corresponding propagator $S_h$ on color indices. Furthermore, in Eq.~\re{F:z},  we do not  display
the  factor $\delta^{(D-1)}(x_{\perp})$.}
\begin{eqnarray}
\Si_h(\mu = 2 \om, \vv =-1) &=&  \Si_h^t + (\delta \Si_h^s)^{\text{Re}} + (\delta \Si_h^s)^{\text{Im}}
\nonumber\\
&=&\sum_{n=1}^3 (\Si^t_h)_n a^n + \sum_{n=1}^3 (\delta \Si_h^s)^{\text{Re}}_n  a^n + i \sum_{n=1}^3 (\delta \Si_h^s)^{\text{Im}}_n a^n\,, 
\end{eqnarray}
where the first term corresponds to the time-like self-energy, 
\beq
\Si_h(\mu = -2 \om, \vv =1)= \Si_h^t =
\sum_{n=1}^3 (\Si^t_h)_n a^n  
\mlabel{Si:t-case}
\,,
\eeq
and the addenda $\delta\Si^s_h$ arise when going over to the space-like case using the substitution rule in Eq.~\re{t->s}, 
\begin{eqnarray}
\ln^n(-2\om) &\to&  \ln^n(2\om)
+ \sum_{j=1}^{j \le [n/2]} \left(\frac{-\pi^2}{4}\right)^j\binom{n}{2j}
 \ln^{n-2j}(2\om)
\nonumber\\
&&{} -i  \sum_{j=0}^{j \le [(n-1)/2]} \frac{\pi}{2}\left(\frac{-\pi^2}{4}\right)^j\binom{n}{2j+1} \ln^{n-2j-1}(2\om)
\,.
\end{eqnarray}
Using Landau gauge, we find the following expressions: 
\begin{eqnarray}
 (\tSih^t)_1&=&
  -\frac{16}{3}\,,
 \qquad 
(\tSih^t)_2=
-\frac{4355}{18}-12 \pi^2+n_f\, \left[\frac{152}{9}+\frac{8}{9} \pi^2\right]
\,, 
\nonumber\\
(\tSih^t)_3&=&
-\frac{3741889}{324}-\frac{8765}{9} \pi^2+\frac{4603}{360} \pi^4+893  \zeta_{3}+44 \pi^2  \zeta_{3}-372  \zeta_{5}
\nonumber\\
&&{}+n_f\, \left[\frac{388024}{243}+\frac{3316}{27} \pi^2-\frac{104}{135} \pi^4-120  \zeta_{3}\right]
+n_f^2\, \left[-\frac{31232}{729}-\frac{256}{81} \pi^2+\frac{160}{27}  \zeta_{3}\right]
\,, 
\nnb
\eea
\myhl
\begin{flalign}
 &(\delta \tSih^s)^{\text{Re}}_1=0, 
 \hspace{4mm} 
(\delta \tSih^s)^{\text{Re}}_2=
14 \pi^2-\frac{4}{3} \pi^2\, n_f
, & \NEQN{deltaSReL1,2}
\\
&(\delta \tSih^s)^{\text{Re}}_3=
\frac{2873}{3} \pi^2 
-\frac{1322}{9} \pi^2\, n_f
+\frac{128}{27} \pi^2\, n_f^2
, & \NEQN{deltaSReL3}
 \end{flalign} 
\vspace{-8mm}
\bea
(\delta \tSih^s)^{\text{Im}}_1&=&
-4 \pi\,, 
\qquad
(\delta \tSih^s)^{\text{Im}}_2=
-97 \pi+\frac{64}{9} \pi\, n_f
\,, 
\nonumber\\
 (\delta \tSih^s)^{\text{Im}}_3&=&
-\frac{188723}{36} \pi-132 \pi^3-\frac{8}{5} \pi^5-\frac{369}{2} \pi  \zeta_{3}
+n_f\, \left[\frac{20236}{27} \pi+\frac{160}{9} \pi^3+\frac{160}{3} \pi  \zeta_{3}\right]
\nonumber\\
&&{}+n_f^2\, \left[-\frac{1744}{81} \pi-\frac{16}{27} \pi^3\right]
\,. \EQN{deltaSImL3}
\end{eqnarray} 
Notice that the full dependence on $\ln[\mu/(2\om)]$ in Eq.~\re{deltaSImL3}
can be easily restored from the evolution Eq.~\re{RG:hself} with the use of the
AD $\ga_h$  as given in Appendix A.
\setlength{\hlength}{-11mm}
 
Assuming $n_f=3$ and substituting $a_s=4 a \equiv \alpha_s/\pi$, we get
numerically:
\begin{eqnarray}
 (\tSih^t)_{n_f=3}&=& -1.33333 \,a_s -17.7121 \, a_s^2 -180.297 \, a_s^3 
  \,,
\nonumber\\
(\delta \tSih^s)_{n_f=3}^{\text{Re}}  &=&6.1685 \, a_s^2 +86.3076 \, a_s^3 
\,,
\nonumber\\
(\delta \tSih^s)_{n_f=3}^{\text{Im}}  &=&-3.14159 \,a_s -14.8571 \, a_s^2 -206.265 \, a_s^3 
\,. \EQN{deltaSImNL:nf3}
 \end{eqnarray} 
Notice that the heavy-field self-energy at $v^2=-1$ acquires a large imaginary part, whereas the difference 
in the real part is minor.

Next, we consider the momentum-space correlation function in Eq.~\re{Pi(omega)}.
Our result for $\tD$ reads\footnote{%
Notice that there is an overall minus sign
between the time-like and space-like correlation functions, unlike for the self-energy $\Sigma_h$.}
\begin{eqnarray}
{\pi^2}\tilde{D}(\mu = 2 \om, \vv =-1) &=& -\tilde{D}^t + (\delta \tilde{D}^s)^{\text{Re}} + (\delta \tilde{D}^s)^{\text{Im}}
\nonumber\\
&=&- \sum_{n=0}^2 (\tilde{D}^t)_n a^n + \sum_{n=0}^2 (\delta \tilde{D}^s)^{\text{Re}}_n  a^n
+ i \sum_{n=0}^2 (\delta \tilde{D}^s)^{\text{Im}}_n a^n 
\mlabel{F:s-case}
\,,
\end{eqnarray}
where, as above, the first term corresponds to the time-like correlation function,
\beq
{\pi^2}\tilde{D}(\mu = -2 \om, \vv =1) = \tilde{D}^t = \sum_{n=0}^2 (\tilde{D}^t)_n a^n
\,.
\mlabel{F:t-case}
\eeq
The coefficients in Eq.~(\ref{F:s-case}) are given by
\begin{eqnarray}
 (	\tD^t)_{0}&=&
3\,,\qquad 
(	\tD^t)_{1}=
68+\frac{16}{3} \pi^2
\,,
\nonumber\\
(	\tD^t)_{2}&=&
\frac{24749}{6}+\frac{3680}{9} \pi^2-\frac{32}{27} \pi^4-\frac{3872}{3}  \zeta_{3}
+n_f\, \Bigl[-\frac{1849}{9}-\frac{328}{27} \pi^2+64  \zeta_{3}\Bigr]
\,,
\nonumber\\
(\delta 	\tD^s)^{\text{Re}}_{0}  &=&0\,, 
 \qquad
 (\delta 	\tD^s)^{\text{Re}}_{1}  =0\,,
\nonumber\\
(\delta 	\tD^s)^{\text{Re}}_{2}  &=&
90 \pi^2 
-4 \pi^2\, n_f\,,
\nonumber\\
(\delta 	\tD^s)^{\text{Im}}_0&=&0\,, 
 \qquad
(\delta 	\tD^s)^{\text{Im}}_1=
-12 \pi\,,
\nonumber\\
(\delta 	\tD^s)^{\text{Im}}_2&=&
-\frac{3314}{3} \pi-\frac{776}{9} \pi^3
+n_f\, \Bigl[52 \pi+\frac{32}{9} \pi^3\Bigr]\,.
\label{deltaDIm2}
 \end{eqnarray} 
For $n_f=3$, we have numerically
\begin{eqnarray}
 (\delta 	\tD^t)_{n_f=3}&=&3.+30.1595 \,a_s +359.267 \, a_s^2\,, 
\nonumber\\
(\delta 	\tD^s)_{n_f=3}^{\text{Re}}  &=&48.1143 \, a_s^2 \,,
\nonumber\\
(\delta 	\tD^s)_{n_f=3}^{\text{Im}}  &=&-9.42478 \,a_s -332.689 \, a_s^2\,.
\EQN{deltaDmNL:nf3}
 \end{eqnarray} 
Finally, we  present below our results for the two momentum-space correlators,
\begin{eqnarray}
\tilde{D}_\TT(\mu = 2 \om, \vv =-1) &=&   a \left(\tilde{D}^t_\TT + 
(\delta \tilde{D}^s_\TT)^{\text{Re}} + (\delta \tilde{D}^s_\TT)^{\text{Im}}
\right)
\nonumber\\
&=& a\Bigl( \sum_{n=0}^2 (\tilde{D}^t_\TT)_n a^n + \sum_{n=0}^3 (\delta \tilde{D}^s_\TT)^{\text{Re}}_n  a^n
+ i \sum_{n=0}^2 (\delta \tilde{D}^s_\TT)^{\text{Im}}_n a^n\Bigr)\,, 
\nonumber\\
\tilde{D}_\PT(\mu = 2 \om, \vv =-1) &=&   a  \left(\tilde{D}^t_\PT + 
(\delta \tilde{D}^s_\PT)^{\text{Re}} + (\delta \tilde{D}^s_\PT)^{\text{Im}}
\right)
\nonumber\\
&& \hspace{-1cm} = a\Bigl(\sum_{n=0}^2 (\tilde{D}^t_\PT)_n a^n + \sum_{n=0}^3 (\delta \tilde{D}^s_\PT)^{\text{Re}}_n  a^n
+ i \sum_{n=0}^2 (\delta \tilde{D}^s_\PT)^{\text{Im}}_n a^n\Bigr), 
\end{eqnarray}
where
\begin{eqnarray}
 (\tD^t_\TT)_{0}&=&
  \frac{64}{3}\,,
  \qquad
(\tD^t_\TT)_{1}=
\frac{8576}{9}+\frac{256}{3} \pi^2-\frac{1664}{27}\, n_f\,,
\nonumber\\
(\tD^t_\TT)_{2}&=&
\frac{1994336}{27}+\frac{94208}{9} \pi^2-19584  \zeta_{3}
+n_f\, \Bigl[-\frac{692224}{81}-\frac{15616}{27} \pi^2+\frac{8960}{9}  \zeta_{3}\Bigr]
\nonumber\\
&&{}+n_f^2\, \Bigl[\frac{53504}{243}+\frac{512}{81} \pi^2\Bigr]\,,
\nonumber\\
(\delta \tD^s_\TT)^{\text{Re}}_{0}  &=&0\,, 
\qquad
(\delta \tD^s_\TT)^{\text{Re}}_{1}  =0\,, 
\nonumber\\
 (\delta \tD^s_\TT)^{\text{Re}}_{2}  &=&
-\frac{2560}{3} \pi^2 
+\frac{1664}{9} \pi^2\, n_f
-\frac{256}{27} \pi^2\, n_f^2\,,
\nonumber\\
 (\delta \tD^s_\TT)^{\text{Im}}_0&=&0\,, 
\qquad
(\delta \tD^s_\TT)^{\text{Im}}_1=
\frac{320}{3} \pi-\frac{128}{9} \pi\, n_f\,,
\nonumber\\
(\delta \tD^s_\TT)^{\text{Im}}_2&=&
\frac{143744}{9} \pi+\frac{4864}{3} \pi^3
+n_f\, \Bigl[-\frac{63232}{27} \pi-\frac{1024}{9} \pi^3\Bigr]
+\frac{6656}{81} \pi\, n_f^2\,,
\nonumber\\
(\tD^t_\PT)_{0}&=&
-\frac{64}{3}\,, 
\qquad
(\tD^t_\PT)_{1}=
-\frac{9536}{9}-\frac{256}{3} \pi^2+\frac{1280}{27}\, n_f\,,
\nonumber\\
(\tD^t_\PT)_{2}&=&
-\frac{2253920}{27}-\frac{97088}{9} \pi^2+21888  \zeta_{3}
+n_f\, \Bigl[\frac{638656}{81}+\frac{14848}{27} \pi^2-\frac{8960}{9}  \zeta_{3}\Bigr]
\nonumber\\
&&{}+n_f^2\, \Bigl[-\frac{33536}{243}-\frac{512}{81} \pi^2\Bigr]\,,
\nonumber\\
(\delta \tD^s_\PT)^{\text{Re}}_{0}  &=&0\,, 
\qquad
(\delta \tD^s_\PT)^{\text{Re}}_{1}  =0\,, 
\nonumber\\
 (\delta \tD^s_\PT)^{\text{Re}}_{2} & =&
\frac{7744}{3} \pi^2 
-\frac{2816}{9} \pi^2\, n_f
+\frac{256}{27} \pi^2\, n_f^2\,,
\nonumber\\
(\delta \tD^s_\PT)^{\text{Im}}_0&=&0\,, 
\qquad
(\delta \tD^s_\PT)^{\text{Im}}_1=
-\frac{704}{3} \pi+\frac{128}{9} \pi\, n_f\,,
\nonumber\\
(\delta \tD^s_\PT)^{\text{Im}}_2&=&
-\frac{229376}{9} \pi-\frac{6400}{3} \pi^3
+n_f\, \Bigl[\frac{73600}{27} \pi+\frac{1024}{9} \pi^3\Bigr]
-\frac{5120}{81} \pi\, n_f^2\,.
\end{eqnarray}
For $n_f=3$, one obtains numerically:
\begin{eqnarray}
 (\delta \tD^t_\TT)_{n_f=3}&=&\frac{64}{3}\Bigl(1+18.8696 \,a_s +342.786 \, a_s^2 \Bigr)\,, 
\nonumber\\
(\delta \tD^s_\TT)_{n_f=3}^{\text{Re}}  &=&-236.871 \, a_s^2\,,
\nonumber\\
(\delta \tD^s_\TT)_{n_f=3}^{\text{Im}}  &=&50.2655 \,a_s +4382.21 \, a_s^2\,, 
\nonumber\\
(\delta \tD^t_\PT)_{n_f=3}&=&-\frac{64}{3}\Bigl(1+20.6196 \,a_s +378.205 \, a_s^2 \Bigr)\,,
\nonumber\\
(\delta \tD^s_\PT)_{n_f=3}^{\text{Re}}  &=&1065.92 \, a_s^2\,,
\nonumber\\
(\delta \tD^s_\PT)_{n_f=3}^{\text{Im}}  &=&-150.796 \,a_s -6982.91 \, a_s^2\,.
\end{eqnarray}

\subsection{\label{corr:x}Correlation functions: position space}

As follows from Eq.~\eqref{sub}, the 
time-like and space-like renormalized correlation functions in position space 
are given by identical expressions with the substitution $it \to z$.
Our result for the ``heavy'' propagator in position space is
in agreement with Eq.~(12) of Ref.~\cite{Chetyrkin:2003vi}
derived for $\vv=1$. 
For the correlation function in Eq.~\eqref{O(z)}, we obtain
\beq
\Pi(z>0, \vv=-1) =  \, \frac{6}{i\,\pi^2\, z^3} \left( 1 +  \sum_{n=1}^3 \left(F\right)_n  a^n\right)
\mlabel{F:z}
\,,
\eeq
with coefficients
\begin{eqnarray}
 (F)_1&=&
\frac{32}{3}+\frac{16}{9} \pi^2+8 L_z  
\,, 
\nonumber\\
 (F)_2&=&
\frac{7025}{18}+\frac{812}{27} \pi^2-\frac{32}{81} \pi^4-\frac{3872}{9}  \zeta_{3}+\frac{3388}{9} L_z  +\frac{1552}{27} \pi^2 L_z  +120 \, L_z^2
\nonumber\\
&&{}+n_f\, \left[-\frac{589}{27}+\frac{32}{81} \pi^2+\frac{64}{3}  \zeta_{3}-\frac{56}{3} L_z  -\frac{64}{27} \pi^2 L_z  -\frac{16}{3} \, L_z^2\right]
\,,  \EQN{F2}
\end{eqnarray} 
where $L_z = \ln\left(  \mu \mathrm{e}^{\g_E}z/2\right)$. 
For $n_f=3$, we get numerically:
\begin{equation}
(F)_{n_f=3}=1.0+7.05316 \,a_s +2.0\, L_z   \,a_s +9.66546 \, a_s^2 +51.0988\, L_z   \, a_s^2 +6.5 \, L_z^2 \, a_s^2 
\,. \EQN{xFN}
\end{equation}
Notice that the higher-order coefficients in position space are considerably smaller than in momentum space, cf. 
Eq.~\re{deltaSImNL:nf3}. \\
Our results for the position-space functions $\Pi_\TT$ and $\Pi_\PT$, 
\begin{eqnarray}
\Pi_\TT(z>0, \vv=-1) &=& \frac{128\,a}{ z^4} \left( 1 +  \sum_{n=1}^2 \left( F_\TT \right)_n  a^n\right)\,,
\nonumber\\
\Pi_\PT(z>0, \vv=-1) &=& -\frac{128\, a}{ z^4} \left( 1 +  \sum_{n=1}^2 \left( F_\PT \right)_n  a^n\right)\,,
\end{eqnarray}
are given by
\begin{eqnarray}
 (F_\TT)_1&=&
\frac{79}{3}+4 \pi^2+10 L_z  
+n_f\, \Bigl[-\frac{4}{9}-\frac{4}{3} L_z  \Bigr]\,,
\nonumber\\
(F_\TT)_2&=&
\frac{18671}{18}+\frac{556}{3} \pi^2-918  \zeta_{3}+\frac{2732}{3} L_z  +152 \pi^2 L_z  +160 \, L_z^2
\nonumber\\
&&{}+n_f\, \Bigl[-\frac{1820}{27}-\frac{16}{9} \pi^2+\frac{140}{3}  \zeta_{3}-\frac{832}{9} L_z  -\frac{32}{3} \pi^2 L_z  -\frac{104}{3} \, L_z^2\Bigr]
\nonumber\\
&&{}+n_f^2\, \Bigl[-\frac{20}{81}+\frac{32}{27} L_z  +\frac{16}{9} \, L_z^2\Bigr]
\,,
\nonumber\\
(F_\PT)_1&=&
\frac{28}{3}+4 \pi^2+22 L_z  
+n_f\, \Bigl[\frac{2}{9}-\frac{4}{3} L_z  \Bigr]\,,
\nonumber\\
(F_\PT)_2&=&
\frac{9011}{18}+\frac{175}{3} \pi^2-1026  \zeta_{3}+\frac{1844}{3} L_z  +200 \pi^2 L_z  +484 \, L_z^2
\nonumber\\
&&{}+n_f\, \Bigl[-\frac{497}{27}+\frac{32}{9} \pi^2+\frac{140}{3}  \zeta_{3}-\frac{364}{9} L_z  -\frac{32}{3} \pi^2 L_z  -\frac{176}{3} \, L_z^2\Bigr]
\nonumber\\
&&{}+n_f^2\, \Bigl[-\frac{68}{81}-\frac{16}{27} L_z  +\frac{16}{9} \, L_z^2\Bigr]\,.
\end{eqnarray}
For $n_f=3$, one obtains numerically:
\begin{eqnarray}
 (F_\TT)_{n_f=3} &=&1+16.1196 \,a_s +1.5 L_z   \,a_s +104.635 \, a_s^2 +114.272 L_z   \, a_s^2 +4.5 \, L_z^2 \, a_s^2\,, 
\nonumber\\
(F_\PT)_{n_f=3} &=&1+12.3696 \,a_s +4.5 L_z   \,a_s +3.36335 \, a_s^2 +134.131 L_z   \, a_s^2 +20.25 \, L_z^2 \, a_s^2\,.
\nonumber\\
&&
\end{eqnarray}

\newcommand{\als}{\left(\frac{\alpha_e}{\pi}\right)}

In the case of QED, the gluon non-local operator in Eq.~\re{F(z)} contracted with $g^{\mu\al}\, g^{\nu\beta}$
can be interpreted as a ``photonic condensate'' regulated with the splitting technique \cite{Vainshtein:1989ve}. 
In terms of the 
scalar functions $\Pi_\TT$ and $\Pi_\PT$, we have
\beq
\F^{\text{QED},\mu\nu}_{\mu\nu}(z) = 6\left(\Pi_\TT + \Pi_\PT\right)= -\frac{1}{8 z^4} \als^2
\left[
1 + \left(\frac{3}{4} -\frac{4}{9}\, n_f - \frac{4}{3}\, n_f\, L_z\right)\frac{\alpha_e}{\pi}
\right]
\,.
\eeq
where $\alpha_e$ is the fine-structure constant.
Notice that, in the QED case, the Wilson line appearing in Eq.~\re{F(z)} then
reduces to just 1. Thus, 
$\F^{\text{QED},\mu\nu}_{\mu\nu}(z)$ is directly expressible in terms of the photon
propagator in position space. The latter is currently known through order
$\alpha_e^5$ \cite{Chetyrkin:2010dx,Baikov:2012zm}.

\section{Conclusions}
\label{sec:six}

We have studied the renormalization and vacuum expectation values of 
non-local off-light-cone operators of a quark and an antiquark field, Eq.~(\ref{O(z)}), and also of two gluon field strength tensors,
Eq.~(\ref{F(z)}), connected by a straight-line-ordered Wilson line, Eq.~(\ref{wline}).

Nucleon matrix elements of these operators are usually called qPDFs and they
are amenable to nonperturbative calculations on the lattice for
space-like separations of the quark fields.   
At the same time, they are counterparts of similar time-like matrix elements that have been 
discussed in the past in the context of heavy-quark expansion in $B$-meson weak decays,
and it is important to understand the relation between time-like and space-like renormalization and matrix elements.

We have shown, to all orders in perturbation theory, that the results for a generic Green function involving a qPDF operator 
at space-like and time-like separations are related by a
specific substitution rule reflecting analytic continuation in the square $\vv$
of the four-vector $v^\mu$ pointing along the Wilson line; see Eq.~(\ref{t->s}).
The RCs and ADs are the same for space-like and time-like separations.
This result is also relevant in the context of
TMD factorization, where Wilson lines are shifted off the light cone to      
regularize rapidity divergences in TMD operators~\cite{Collins:2011zzd}. Our statement is that
the ADs and RCs do not depend on the direction of the shift, space-like or time-like. 

We have calculated the self-energy of the ``heavy'' field $h_v$ in the effective field theory of Eq.~(\ref{lgr}), 
the quark-antiquark qPDF AD and VEV,  
and all the RCs and ADs that are involved in the respective renormalization
through three loops in the $\overline{\mbox{MS}}$ scheme.  Our results
agree with the literature as far as it goes.  
In addition, we have clarified the general RG pattern for the gluon qPDF operator in Eq.~(\ref{F(z)})
and calculated its VEV through three loops, from which the two-loop ADs can be extracted avoiding pollution by gauge-noninvariant operators.

Our results can be used in lattice calculations aiming at the determination of quark and gluon PDFs, e.g., in the nucleon, if the linear UV divergences of lattice observables are removed by considering suitable ratios of matrix
elements involving the same operator \cite{Orginos:2017kos,Braun:2018brg}.

\section*{Acknowledgments}

We thank Andrey Grozin for reading  the  manuscript and  valuable advice. 
The work of V.M.B. and B.A.K. was supported in part by the DFG 
Research Unit FOR 2926 under Grant No.\ 409651613.
The work of K.G.C. was supported in part by DFG grant CH~1479/2-1. 
%
\appendix
\section{Anomalous dimensions}
\label{app:a}
Representing a generic AD $\ga(a) $ as
\beq
\ga = \sum_{n \ge 1}
\left(\ga\right)_n a^n
\,,
\eeq
we have for the coefficients relevant here 
\begin{eqnarray}
(\beta)_1&=&
  -11+\frac{2}{3}\, n_f\,,
 \qquad
(\beta)_2=
  -102+\frac{38}{3}\, n_f\,,
\nonumber\\
 (\beta)_3&=&
-\frac{2857}{2} 
+\frac{5033}{18}\, n_f
-\frac{325}{54}\, n_f^2
\,, 
\nonumber\\
(\ga_Q)_1&=&
2\,, 
\qquad
(\ga_Q)_2=
\frac{127}{9}+\frac{28}{27} \pi^2-\frac{10}{9}\, n_f
\,, 
\nonumber\\
(\ga_Q)_3&=&
-\frac{61}{6}+\frac{1372}{81} \pi^2+\frac{760}{243} \pi^4-\frac{356}{9}  \zeta_{3}
+n_f\, \left[-\frac{344}{81}-\frac{392}{243} \pi^2-\frac{664}{27}  \zeta_{3}\right]
-\frac{70}{81}\, n_f^2
\,, 
\nonumber
\\
(\ga_h)_1&=&
4\,, 
\qquad
(\ga_h)_2=
\frac{179}{3}-\frac{32}{9}\, n_f
\,, 
\nonumber\\
(\ga_h)_3&=&
\frac{23815}{36}+\frac{8}{5} \pi^4+\frac{369}{2}  \zeta_{3}
+n_f\, \left[-\frac{2006}{27}-\frac{160}{3}  \zeta_{3}\right]
-\frac{80}{81}\, n_f^2
\,, 
\nonumber\\
(\ga_2)_1&=&0
\,, 
\qquad
(\ga_2)_2=
-\frac{67}{3}+\frac{4}{3}\, n_f
\,, 
\nonumber\\
(\ga_2)_3&=&
-\frac{20729}{36}+\frac{79}{2}  \zeta_{3} 
+\frac{550}{9}\, n_f
-\frac{20}{27}\, n_f^2
\,, 
\nonumber\\
(\ga_{\perp\perp})_1 &=& -3
\,,
\qquad
(\ga_{\perp\perp})_2 = -34 +6\, \pi^2 +\frac{13}{3}\, n_f
\,,
\nonumber\\
(\ga_{\parallel\perp})_1 &=& 0
\,,
\qquad
(\ga_{\parallel\perp})_2 =  6\, \pi^2 
\,.
\end{eqnarray} 

\section{Renormalization constants}
\label{app:b}

Representing a generic RC $Z(a,\ep) $ as
\beq
Z(a,\ep) =1+\sum_{n \ge  1}
\left( Z(\ep) \right)_n \left(\frac{a}{\ep}\right)^n
\,,
\eeq
we have for the coefficients relevant here
\begin{eqnarray}
(Z_a)_1&=&
-11+\frac{2}{3}\, n_f\,, 
\qquad
(Z_a)_2=
121-51 \ep+n_f\, \left[-\frac{44}{3}+\frac{19}{3} \ep\right]+\frac{4}{9}\, n_f^2
\,, 
\nonumber\\
(Z_a)_3&=&
-1331+1309 \ep-\frac{2857}{6} \ep^2
+n_f\, \left[242-\frac{2177}{9} \ep+\frac{5033}{54} \ep^2\right]
\nonumber\\
&&{}+n_f^2\, \left[-\frac{44}{3}+\frac{266}{27} \ep-\frac{325}{162} \ep^2\right]+\frac{8}{27}\, n_f^3
\,,  
\nonumber\\
(Z_Q)_1&=&
-2\,,
\qquad
(Z_Q)_2=
13-\frac{127}{18} \ep-\frac{14}{27} \pi^2 \ep+n_f\, \left[-\frac{2}{3}+\frac{5}{9} \ep\right]
\,, 
\nonumber\\
(Z_Q)_3&=&
-104+\frac{3614}{27} \ep+\frac{392}{81} \pi^2 \ep+\frac{61}{18} \ep^2-\frac{1372}{243} \pi^2 \ep^2-\frac{760}{729} \pi^4 \ep^2+\frac{356}{27}  \zeta_{3} \ep^2
\nonumber\\
&&{}+n_f\, \left[\frac{100}{9}-\frac{1358}{81} \ep-\frac{56}{243} \pi^2 \ep+\frac{344}{243} \ep^2+\frac{392}{729} \pi^2 \ep^2+\frac{664}{81}  \zeta_{3} \ep^2\right]
\nonumber\\
&&{}+n_f^2\, \left[-\frac{8}{27}+\frac{20}{81} \ep+\frac{70}{243} \ep^2\right]
\,, 
\nonumber\\
(Z_h)_1&=&
4\,, 
\qquad
(Z_h)_2=
-14+\frac{179}{6} \ep+n_f\, \left[\frac{4}{3}-\frac{16}{9} \ep\right]
\,,  
\nonumber\\
(Z_h)_3&=&
84-\frac{2119}{9} \ep+\frac{23815}{108} \ep^2+\frac{8}{15} \pi^4 \ep^2+\frac{123}{2}  \zeta_{3} \ep^2
\nonumber\\
&&{}+n_f\, \left[-\frac{128}{9}+\frac{974}{27} \ep-\frac{2006}{81} \ep^2-\frac{160}{9}  \zeta_{3} \ep^2\right]
+n_f^2\, \left[\frac{16}{27}-\frac{64}{81} \ep-\frac{80}{243} \ep^2\right]
\,, 
\nonumber\\
(Z_2)_1&=0&
\,, 
\qquad
(Z_2)_2=
-\frac{67}{6} \ep+\frac{2}{3} \ep\, n_f
\,, 
\nonumber\\
(Z_2)_3&=&
\frac{737}{9} \ep-\frac{20729}{108} \ep^2+\frac{79}{6}  \zeta_{3} \ep^2
+n_f\, \left[-\frac{266}{27} \ep+\frac{550}{27} \ep^2\right]
+n_f^2\, \left[\frac{8}{27} \ep-\frac{20}{81} \ep^2\right]
\,.\qquad 
\nonumber\\
(Z_{\perp\perp})_1 &=& 3
\,, 
\qquad
(Z_{\perp\perp})_2 = -12 + (17 - 3\,\pi^2)\,\ep +  n_f 
\left[ 1 -\frac{13}{6}\,\ep \right]
\,, 
\nonumber\\
(Z_{\parallel\perp})_1  &=&  0,
\qquad
(Z_{\parallel\perp})_2   =  -3\,  \pi^2 \ep
\,.
\end{eqnarray} 

\begin{thebibliography}{10}

\bibitem{Makeenko:1980vm}
{\relax Yu}.~Makeenko and A.~A. Migdal, \emph{{Quantum Chromodynamics as
  Dynamics of Loops}},
  \href{http://dx.doi.org/10.1016/0550-3213(81)90258-3}{\emph{Nucl. Phys.}
  {\bfseries B188} (1981) 269}.

\bibitem{Polyakov:1980ca}
A.~M. Polyakov, \emph{{Gauge Fields as Rings of Glue}},
  \href{http://dx.doi.org/10.1016/0550-3213(80)90507-6}{\emph{Nucl. Phys.}
  {\bfseries B164} (1980) 171--188}.

\bibitem{Gervais:1979fv}
J.-L. Gervais and A.~Neveu, \emph{{The Slope of the Leading Regge Trajectory in
  Quantum Chromodynamics}},
  \href{http://dx.doi.org/10.1016/0550-3213(80)90397-1}{\emph{Nucl. Phys.}
  {\bfseries B163} (1980) 189--216}.

\bibitem{Dotsenko:1979wb}
V.~S. Dotsenko and S.~N. Vergeles, \emph{{Renormalizability of Phase Factors in
  the Nonabelian Gauge Theory}},
  \href{http://dx.doi.org/10.1016/0550-3213(80)90103-0}{\emph{Nucl. Phys.}
  {\bfseries B169} (1980) 527--546}.

\bibitem{Craigie:1980qs}
N.~S. Craigie and H.~Dorn, \emph{{On the Renormalization and Short Distance
  Properties of Hadronic Operators in {QCD}}},
  \href{http://dx.doi.org/10.1016/0550-3213(81)90372-2}{\emph{Nucl. Phys.}
  {\bfseries B185} (1981) 204--220}.

\bibitem{Arefeva:1980zd}
I.~{\relax Ya}. Arefeva, \emph{{Quantum contour field equations}},
  \href{http://dx.doi.org/10.1016/0370-2693(80)90529-8}{\emph{Phys. Lett.}
  {\bfseries 93B} (1980) 347--353}.

\bibitem{Brandt:1981kf}
R.~A. Brandt, F.~Neri and M.-a. Sato, \emph{{Renormalization of Loop Functions
  for All Loops}}, \href{http://dx.doi.org/10.1103/PhysRevD.24.879}{\emph{Phys.
  Rev.} {\bfseries D24} (1981) 879}.

\bibitem{Dorn:1986dt}
H.~Dorn, \emph{{Renormalization of Path Ordered Phase Factors and Related
  Hadron Operators in Gauge Field Theories}},
  \href{http://dx.doi.org/10.1002/prop.19860340104}{\emph{Fortsch. Phys.}
  {\bfseries 34} (1986) 11--56}.

\bibitem{Korchemsky:1985xj}
G.~P. Korchemsky and A.~V. Radyushkin, \emph{{Loop Space Formalism and
  Renormalization Group for the Infrared Asymptotics of {QCD}}},
  \href{http://dx.doi.org/10.1016/0370-2693(86)91439-5}{\emph{Phys. Lett.}
  {\bfseries B171} (1986) 459--467}.

\bibitem{Korchemsky:1987wg}
G.~P. Korchemsky and A.~V. Radyushkin, \emph{{Renormalization of the Wilson
  Loops Beyond the Leading Order}},
  \href{http://dx.doi.org/10.1016/0550-3213(87)90277-X}{\emph{Nucl. Phys.}
  {\bfseries B283} (1987) 342--364}.

\bibitem{Korchemsky:1991zp}
G.~P. Korchemsky and A.~V. Radyushkin, \emph{{Infrared factorization, Wilson
  lines and the heavy quark limit}},
  \href{http://dx.doi.org/10.1016/0370-2693(92)90405-S}{\emph{Phys. Lett.}
  {\bfseries B279} (1992) 359--366},
  [\href{https://arxiv.org/abs/hep-ph/9203222}{{\ttfamily hep-ph/9203222}}].

\bibitem{Broadhurst:1991fz}
D.~J. Broadhurst and A.~G. Grozin, \emph{{Two loop renormalization of the
  effective field theory of a static quark}},
  \href{http://dx.doi.org/10.1016/0370-2693(91)90532-U}{\emph{Phys. Lett.}
  {\bfseries B267} (1991) 105--110},
  [\href{https://arxiv.org/abs/hep-ph/9908362}{{\ttfamily hep-ph/9908362}}].

\bibitem{Chetyrkin:2003vi}
K.~G. Chetyrkin and A.~G. Grozin, \emph{{Three loop anomalous dimension of the
  heavy light quark current in HQET}},
  \href{http://dx.doi.org/10.1016/S0550-3213(03)00490-5}{\emph{Nucl. Phys.}
  {\bfseries B666} (2003) 289--302},
  [\href{https://arxiv.org/abs/hep-ph/0303113}{{\ttfamily hep-ph/0303113}}].

\bibitem{Grozin:2007ap}
A.~G. Grozin, T.~Huber and D.~Maitre, \emph{{On one master integral for
  three-loop on-shell HQET propagator diagrams with mass}},
  \href{http://dx.doi.org/10.1088/1126-6708/2007/07/033}{\emph{JHEP} {\bfseries
  07} (2007) 033}, [\href{https://arxiv.org/abs/0705.2609}{{\ttfamily
  0705.2609}}].

\bibitem{Grozin:2008nu}
A.~G. Grozin and R.~N. Lee, \emph{{Three-loop HQET vertex diagrams for B0 -
  anti-B0 mixing}},
  \href{http://dx.doi.org/10.1088/1126-6708/2009/02/047}{\emph{JHEP} {\bfseries
  02} (2009) 047}, [\href{https://arxiv.org/abs/0812.4522}{{\ttfamily
  0812.4522}}].

\bibitem{Collins:2011zzd}
J.~Collins, \emph{{Foundations of perturbative QCD}}, {\emph{Camb. Monogr.
  Part. Phys. Nucl. Phys. Cosmol.} {\bfseries 32} (2011) 1--624}.

\bibitem{Ji:2013dva}
X.~Ji, \emph{{Parton Physics on a Euclidean Lattice}},
  \href{http://dx.doi.org/10.1103/PhysRevLett.110.262002}{\emph{Phys. Rev.
  Lett.} {\bfseries 110} (2013) 262002},
  [\href{https://arxiv.org/abs/1305.1539}{{\ttfamily 1305.1539}}].

\bibitem{Radyushkin:2017cyf}
A.~V. Radyushkin, \emph{{Quasi-parton distribution functions, momentum
  distributions, and pseudo-parton distribution functions}},
  \href{http://dx.doi.org/10.1103/PhysRevD.96.034025}{\emph{Phys. Rev.}
  {\bfseries D96} (2017) 034025},
  [\href{https://arxiv.org/abs/1705.01488}{{\ttfamily 1705.01488}}].

\bibitem{Izubuchi:2018srq}
T.~Izubuchi, X.~Ji, L.~Jin, I.~W. Stewart and Y.~Zhao, \emph{{Factorization
  Theorem Relating Euclidean and Light-Cone Parton Distributions}},
  \href{http://dx.doi.org/10.1103/PhysRevD.98.056004}{\emph{Phys. Rev.}
  {\bfseries D98} (2018) 056004},
  [\href{https://arxiv.org/abs/1801.03917}{{\ttfamily 1801.03917}}].

\bibitem{Lin:2017snn}
H.-W. Lin et~al., \emph{{Parton distributions and lattice QCD calculations: a
  community white paper}},
  \href{http://dx.doi.org/10.1016/j.ppnp.2018.01.007}{\emph{Prog. Part. Nucl.
  Phys.} {\bfseries 100} (2018) 107--160},
  [\href{https://arxiv.org/abs/1711.07916}{{\ttfamily 1711.07916}}].

\bibitem{Cichy:2018mum}
K.~Cichy and M.~Constantinou, \emph{{A guide to light-cone PDFs from Lattice
  QCD: an overview of approaches, techniques and results}},
  \href{http://dx.doi.org/10.1155/2019/3036904}{\emph{Adv. High Energy Phys.}
  {\bfseries 2019} (2019) 3036904},
  [\href{https://arxiv.org/abs/1811.07248}{{\ttfamily 1811.07248}}].

\bibitem{Detmold:2005gg}
W.~Detmold and C.~J.~D. Lin, \emph{{Deep-inelastic scattering and the operator
  product expansion in lattice QCD}},
  \href{http://dx.doi.org/10.1103/PhysRevD.73.014501}{\emph{Phys. Rev.}
  {\bfseries D73} (2006) 014501},
  [\href{https://arxiv.org/abs/hep-lat/0507007}{{\ttfamily hep-lat/0507007}}].

\bibitem{Braun:2007wv}
V.~M. Braun and D.~M{\"u}ller, \emph{{Exclusive processes in position space and
  the pion distribution amplitude}},
  \href{http://dx.doi.org/10.1140/epjc/s10052-008-0608-4}{\emph{Eur. Phys. J.}
  {\bfseries C55} (2008) 349--361},
  [\href{https://arxiv.org/abs/0709.1348}{{\ttfamily 0709.1348}}].

\bibitem{Ma:2017pxb}
Y.-Q. Ma and J.-W. Qiu, \emph{{Exploring Partonic Structure of Hadrons Using ab
  initio Lattice QCD Calculations}},
  \href{http://dx.doi.org/10.1103/PhysRevLett.120.022003}{\emph{Phys. Rev.
  Lett.} {\bfseries 120} (2018) 022003},
  [\href{https://arxiv.org/abs/1709.03018}{{\ttfamily 1709.03018}}].

\bibitem{Xiong:2013bka}
X.~Xiong, X.~Ji, J.-H. Zhang and Y.~Zhao, \emph{{One-loop matching for parton
  distributions: Nonsinglet case}},
  \href{http://dx.doi.org/10.1103/PhysRevD.90.014051}{\emph{Phys. Rev.}
  {\bfseries D90} (2014) 014051},
  [\href{https://arxiv.org/abs/1310.7471}{{\ttfamily 1310.7471}}].

\bibitem{Ji:2015jwa}
X.~Ji and J.-H. Zhang, \emph{{Renormalization of quasiparton distribution}},
  \href{http://dx.doi.org/10.1103/PhysRevD.92.034006}{\emph{Phys. Rev.}
  {\bfseries D92} (2015) 034006},
  [\href{https://arxiv.org/abs/1505.07699}{{\ttfamily 1505.07699}}].

\bibitem{Ji:2017oey}
X.~Ji, J.-H. Zhang and Y.~Zhao, \emph{{Renormalization in Large Momentum
  Effective Theory of Parton Physics}},
  \href{http://dx.doi.org/10.1103/PhysRevLett.120.112001}{\emph{Phys. Rev.
  Lett.} {\bfseries 120} (2018) 112001},
  [\href{https://arxiv.org/abs/1706.08962}{{\ttfamily 1706.08962}}].

\bibitem{Ishikawa:2017faj}
T.~Ishikawa, Y.-Q. Ma, J.-W. Qiu and S.~Yoshida, \emph{{Renormalizability of
  quasiparton distribution functions}},
  \href{http://dx.doi.org/10.1103/PhysRevD.96.094019}{\emph{Phys. Rev.}
  {\bfseries D96} (2017) 094019},
  [\href{https://arxiv.org/abs/1707.03107}{{\ttfamily 1707.03107}}].

\bibitem{Wang:2017eel}
W.~Wang and S.~Zhao, \emph{{On the power divergence in quasi gluon distribution
  function}}, \href{http://dx.doi.org/10.1007/JHEP05(2018)142}{\emph{JHEP}
  {\bfseries 05} (2018) 142},
  [\href{https://arxiv.org/abs/1712.09247}{{\ttfamily 1712.09247}}].

\bibitem{Wang:2017qyg}
W.~Wang, S.~Zhao and R.~Zhu, \emph{{Gluon quasidistribution function at one
  loop}}, \href{http://dx.doi.org/10.1140/epjc/s10052-018-5617-3}{\emph{Eur.
  Phys. J.} {\bfseries C78} (2018) 147},
  [\href{https://arxiv.org/abs/1708.02458}{{\ttfamily 1708.02458}}].

\bibitem{Zhang:2018diq}
J.-H. Zhang, X.~Ji, A.~Sch{\"a}fer, W.~Wang and S.~Zhao, \emph{{Accessing Gluon
  Parton Distributions in Large Momentum Effective Theory}},
  \href{http://dx.doi.org/10.1103/PhysRevLett.122.142001}{\emph{Phys. Rev.
  Lett.} {\bfseries 122} (2019) 142001},
  [\href{https://arxiv.org/abs/1808.10824}{{\ttfamily 1808.10824}}].

\bibitem{Wang:2019tgg}
W.~Wang, J.-H. Zhang, S.~Zhao and R.~Zhu, \emph{{Complete matching for
  quasidistribution functions in large momentum effective theory}},
  \href{http://dx.doi.org/10.1103/PhysRevD.100.074509}{\emph{Phys. Rev.}
  {\bfseries D100} (2019) 074509},
  [\href{https://arxiv.org/abs/1904.00978}{{\ttfamily 1904.00978}}].

\bibitem{Balitsky:2019krf}
I.~Balitsky, W.~Morris and A.~Radyushkin, \emph{{Gluon Pseudo-Distributions at
  Short Distances: Forward Case}},
  \href{https://arxiv.org/abs/1910.13963}{{\ttfamily 1910.13963}}.

\bibitem{Orginos:2017kos}
K.~Orginos, A.~Radyushkin, J.~Karpie and S.~Zafeiropoulos, \emph{{Lattice QCD
  exploration of parton pseudo-distribution functions}},
  \href{http://dx.doi.org/10.1103/PhysRevD.96.094503}{\emph{Phys. Rev.}
  {\bfseries D96} (2017) 094503},
  [\href{https://arxiv.org/abs/1706.05373}{{\ttfamily 1706.05373}}].

\bibitem{Braun:2018brg}
V.~M. Braun, A.~Vladimirov and J.-H. Zhang, \emph{{Power corrections and
  renormalons in parton quasidistributions}},
  \href{http://dx.doi.org/10.1103/PhysRevD.99.014013}{\emph{Phys. Rev.}
  {\bfseries D99} (2019) 014013},
  [\href{https://arxiv.org/abs/1810.00048}{{\ttfamily 1810.00048}}].

\bibitem{DiGiacomo:2000irz}
A.~Di~Giacomo, H.~G. Dosch, V.~I. Shevchenko and {\relax Yu}.~A. Simonov,
  \emph{{Field correlators in QCD: Theory and applications}},
  \href{http://dx.doi.org/10.1016/S0370-1573(02)00140-0}{\emph{Phys. Rept.}
  {\bfseries 372} (2002) 319--368},
  [\href{https://arxiv.org/abs/hep-ph/0007223}{{\ttfamily hep-ph/0007223}}].

\bibitem{Dorn:1980hs}
H.~Dorn and E.~Wieczorek, \emph{{Renormalization and Short Distance Properties
  of String Type Equations in {QCD}}},
  \href{http://dx.doi.org/10.1007/BF01554111}{\emph{Z. Phys.} {\bfseries C9}
  (1981) 49}.

\bibitem{Dorn:1981wa}
H.~Dorn, D.~Robaschik and E.~Wieczorek, \emph{{Renormalization and short
  distance properties of gauge invariant gluoinum and hadron operators}},
  \href{http://dx.doi.org/10.1002/andp.19834950208}{\emph{Annalen Phys.}
  {\bfseries 40} (1983) 166}.

\bibitem{Neubert:1993mb}
M.~Neubert, \emph{{Heavy quark symmetry}},
  \href{http://dx.doi.org/10.1016/0370-1573(94)90091-4}{\emph{Phys. Rept.}
  {\bfseries 245} (1994) 259--396},
  [\href{https://arxiv.org/abs/hep-ph/9306320}{{\ttfamily hep-ph/9306320}}].

\bibitem{Cicuta:1972jf}
G.~M. Cicuta and E.~Montaldi, \emph{{Analytic renormalization via continuous
  space dimension}}, \href{http://dx.doi.org/10.1007/BF02756527}{\emph{Lett.
  Nuovo Cim.} {\bfseries 4} (1972) 329--332}.

\bibitem{Ashmore:1972uj}
J.~F. Ashmore, \emph{{A Method of Gauge Invariant Regularization}},
  \href{http://dx.doi.org/10.1007/BF02824407}{\emph{Lett. Nuovo Cim.}
  {\bfseries 4} (1972) 289--290}.

\bibitem{tHooft:1972fi}
G.~'t~Hooft and M.~J.~G. Veltman, \emph{{Regularization and Renormalization of
  Gauge Fields}},
  \href{http://dx.doi.org/10.1016/0550-3213(72)90279-9}{\emph{Nucl. Phys.}
  {\bfseries B44} (1972) 189--213}.

\bibitem{Tarasov:1980au}
O.~V. Tarasov, A.~A. Vladimirov and A.~Y. Zharkov, \emph{{The Gell-Mann-Low
  function of QCD in the three loop approximation}}, {\emph{Phys. Lett.}
  {\bfseries B93} (1980) 429--432}.

\bibitem{Tarasov:2019rwk}
O.~V. Tarasov, \emph{{Anomalous dimensions of quark masses in the three-loop
  approximation}},  \href{https://arxiv.org/abs/1910.12231}{{\ttfamily
  1910.12231}}.

\bibitem{Larin:1993tp}
S.~A. Larin and J.~A.~M. Vermaseren, \emph{{The Three loop QCD Beta function
  and anomalous dimensions}},
  \href{http://dx.doi.org/10.1016/0370-2693(93)91441-O}{\emph{Phys. Lett.}
  {\bfseries B303} (1993) 334--336},
  [\href{https://arxiv.org/abs/hep-ph/9302208}{{\ttfamily hep-ph/9302208}}].

\bibitem{Grozin:2004yc}
A.~G. Grozin, \emph{{Heavy quark effective theory}},
  \href{http://dx.doi.org/10.1007/b79301}{\emph{Springer Tracts Mod. Phys.}
  {\bfseries 201} (2004) 1--213}.

\bibitem{QGRAF}
P.~Nogueira, \emph{{Automatic Feynman graph generation}},
  \href{http://dx.doi.org/10.1006/jcph.1993.1074}{\emph{J. Comput. Phys.}
  {\bfseries 105} (1993) 279--289}.

\bibitem{Smirnov:2019qkx}
A.~V. Smirnov and F.~S. Chuharev, \emph{{FIRE6: Feynman Integral REduction with
  Modular Arithmetic}},  \href{https://arxiv.org/abs/1901.07808}{{\ttfamily
  1901.07808}}.

\bibitem{Lee:2012cn}
R.~N. Lee, \emph{{Presenting LiteRed: a tool for the Loop InTEgrals
  REDuction}},  \href{https://arxiv.org/abs/1212.2685}{{\ttfamily 1212.2685}}.

\bibitem{Lee:2013mka}
R.~N. Lee, \emph{{LiteRed 1.4: a powerful tool for reduction of multiloop
  integrals}}, \href{http://dx.doi.org/10.1088/1742-6596/523/1/012059}{\emph{J.
  Phys. Conf. Ser.} {\bfseries 523} (2014) 012059},
  [\href{https://arxiv.org/abs/1310.1145}{{\ttfamily 1310.1145}}].

\bibitem{Grozin:2000jv}
A.~G. Grozin, \emph{{Calculating three loop diagrams in heavy quark effective
  theory with integration by parts recurrence relations}},
  \href{http://dx.doi.org/10.1088/1126-6708/2000/03/013}{\emph{JHEP} {\bfseries
  03} (2000) 013}, [\href{https://arxiv.org/abs/hep-ph/0002266}{{\ttfamily
  hep-ph/0002266}}].

\bibitem{Grozin:2003eg}
A.~G. Grozin, \emph{{Perturbative HQET}},  in \emph{{Heavy quark physics:
  Proceedings, International School, Dubna, Russia, May 27-June 5, 2002}},
  2003.
\newblock \href{https://arxiv.org/abs/hep-ph/0307220}{{\ttfamily
  hep-ph/0307220}}.

\bibitem{Smirnov:2012gma}
V.~A. Smirnov, \emph{{Analytic tools for Feynman integrals}},
  \href{http://dx.doi.org/10.1007/978-3-642-34886-0}{\emph{Springer Tracts Mod.
  Phys.} {\bfseries 250} (2012) 1--296}.

\bibitem{Politzer:1988wp}
H.~D. Politzer and M.~B. Wise, \emph{{Leading Logarithms of Heavy Quark Masses
  in Processes with Light and Heavy Quarks}},
  \href{http://dx.doi.org/10.1016/0370-2693(88)90718-6}{\emph{Phys. Lett.}
  {\bfseries B206} (1988) 681--684}.

\bibitem{Shifman:1986sm}
M.~A. Shifman and M.~B. Voloshin, \emph{{On Annihilation of Mesons Built from
  Heavy and Light Quark and anti-B0 $<\!\!\!-\!\!-\!\!\!>$ B0 Oscillations}},
  {\emph{Sov. J. Nucl. Phys.} {\bfseries 45} (1987) 292}.

\bibitem{Ji:1991pr}
X.-D. Ji and M.~J. Musolf, \emph{{Subleading logarithmic mass dependence in
  heavy meson form-factors}},
  \href{http://dx.doi.org/10.1016/0370-2693(91)91916-J}{\emph{Phys. Lett.}
  {\bfseries B257} (1991) 409--413}.

\bibitem{Melnikov:2000zc}
K.~Melnikov and T.~van Ritbergen, \emph{{The Three loop on-shell
  renormalization of QCD and QED}},
  \href{http://dx.doi.org/10.1016/S0550-3213(00)00526-5}{\emph{Nucl. Phys.}
  {\bfseries B591} (2000) 515--546},
  [\href{https://arxiv.org/abs/hep-ph/0005131}{{\ttfamily hep-ph/0005131}}].

\bibitem{Marquard:2018rwx}
P.~Marquard, A.~V. Smirnov, V.~A. Smirnov and M.~Steinhauser, \emph{{Four-loop
  wave function renormalization in QCD and QED}},
  \href{http://dx.doi.org/10.1103/PhysRevD.97.054032}{\emph{Phys. Rev.}
  {\bfseries D97} (2018) 054032},
  [\href{https://arxiv.org/abs/1801.08292}{{\ttfamily 1801.08292}}].

\bibitem{Bruser:2019auj}
R.~Br{\"u}ser, A.~Grozin, J.~M. Henn and M.~Stahlhofen, \emph{{Matter
  dependence of the four-loop QCD cusp anomalous dimension: from small angles
  to all angles}}, \href{http://dx.doi.org/10.1007/JHEP05(2019)186}{\emph{JHEP}
  {\bfseries 05} (2019) 186},
  [\href{https://arxiv.org/abs/1902.05076}{{\ttfamily 1902.05076}}].

\bibitem{Broadhurst:1991fc}
D.~J. Broadhurst and A.~G. Grozin, \emph{{Operator product expansion in static
  quark effective field theory: Large perturbative correction}},
  \href{http://dx.doi.org/10.1016/0370-2693(92)92009-6}{\emph{Phys. Lett.}
  {\bfseries B274} (1992) 421--427},
  [\href{https://arxiv.org/abs/hep-ph/9908363}{{\ttfamily hep-ph/9908363}}].

\bibitem{Czarnecki:2001rh}
A.~Czarnecki and K.~Melnikov, \emph{{Threshold expansion for heavy light
  systems and flavor off diagonal current current correlators}},
  \href{http://dx.doi.org/10.1103/PhysRevD.66.011502}{\emph{Phys. Rev.}
  {\bfseries D66} (2002) 011502(R)},
  [\href{https://arxiv.org/abs/hep-ph/0110028}{{\ttfamily hep-ph/0110028}}].

\bibitem{Eidemuller:1997bb}
M.~Eidemuller and M.~Jamin, \emph{{QCD field strength correlator at the
  next-to-leading order}},
  \href{http://dx.doi.org/10.1016/S0370-2693(97)01352-X}{\emph{Phys. Lett.}
  {\bfseries B416} (1998) 415--420},
  [\href{https://arxiv.org/abs/hep-ph/9709419}{{\ttfamily hep-ph/9709419}}].

\bibitem{Vainshtein:1989ve}
A.~I. Vainshtein and V.~I. Zakharov, \emph{{Calculating Photonic Condensate in
  Perturbative {QED}}},
  \href{http://dx.doi.org/10.1016/0370-2693(89)90593-5}{\emph{Phys. Lett.}
  {\bfseries B225} (1989) 415--418}.

\bibitem{Chetyrkin:2010dx}
K.~G. Chetyrkin and A.~Maier, \emph{{Massless correlators of vector, scalar and
  tensor currents in position space at orders $\alpha_s^3$ and $\alpha_s^4$:
  Explicit analytical results}},
  \href{http://dx.doi.org/10.1016/j.nuclphysb.2010.11.007}{\emph{Nucl. Phys.}
  {\bfseries B844} (2011) 266--288},
  [\href{https://arxiv.org/abs/1010.1145}{{\ttfamily 1010.1145}}].

\bibitem{Baikov:2012zm}
P.~Baikov, K.~Chetyrkin, J.~K{\"u}hn and J.~Rittinger, \emph{{Vector Correlator
  in Massless QCD at Order ${\cal O}(\alpha_s^4)$ and the QED $\beta$-function
  at Five Loop}}, \href{http://dx.doi.org/10.1007/JHEP07(2012)017}{\emph{JHEP}
  {\bfseries 1207} (2012) 017},
  [\href{https://arxiv.org/abs/1206.1284}{{\ttfamily 1206.1284}}].

\end{thebibliography}

\providecommand{\href}[2]{#2}\begingroup\raggedright\endgroup

\end{document}